\documentclass[fontsize=11pt,a4paper,toc=flat]{arxiv-article}
\bibliography{bibi,non-arxiv}

\usepackage{Definitions}

\usepackage{hyperref}
\graphicspath{{./plots/}{./FIGS/}}

\preprint{\texttt{NT@UW-24-12\\YITP-24-173}}

\def\nab{\overrightarrow{\nabla}}
\def\nabl{\overleftarrow{\nabla}}
\def\edens{{\cal E}}
\def\kf{k_F}
\def\llra{{\relbar\joinrel\longrightarrow}}
\def\mapright#1{{\smash{\mathop{\llra}\limits_{#1}}}}

\DeclareDocumentCommand\Op{ m m}{
  \mathinner{\mathcal{O}_{#1,#2}}
}

\newcommand{\OfficialTitle}{
	Unnuclear matter at large-charge 
}
\title{\setstretch{1.4}
	{\color{Thoughtless}\textls[-20]{\OfficialTitle}}
}

\hypersetup{pdfauthor={Silas R.~Beane, Domenico Orlando, Susanne Reffert},pdftitle={\OfficialTitle},%
	colorlinks=true,linkcolor=ThoughtYouWere,citecolor=ThoughtYouWere,urlcolor=ThoughtYouWere}

\author{%
	\begin{minipage}{.94\textwidth}
		\begin{center} \dosserif%
			{\small
				\textbf{Silas R.~Beane}\textsuperscript{\ding{73}}\textsuperscript{\ding{74}},
				\textbf{Domenico Orlando}\textsuperscript{\ding{72}\ding{73}\ding{71}}, and
  				\textbf{Susanne Reffert}\textsuperscript{\ding{73}\ding{71}} 
			}
		\end{center}
		\authorBlock{\ding{73}}{\dosserif{}%
			Albert Einstein Center for Fundamental Physics,\\
			Institute for Theoretical Physics, University of Bern,\\
			Sidlerstrasse 5, CH-3012 Bern, Switzerland}
  		\authorBlock{\ding{74}}{\dosserif{}%
                        Department of Physics,\\
                        University of Washington,\\
                        Seattle, WA 98195}
		\authorBlock{\ding{72}}{\dosserif{}%
			INFN sezione di Torino.\\
			via Pietro Giuria 1, 10125 Torino, Italy}
       \authorBlock{\ding{71}}{\dosserif{}%
         Yukawa Institute for Theoretical Physics, Kyoto University,\\
         Kyoto 650-0047, Japan}
	\end{minipage}
}

\date{}

\begin{document}

\numberwithin{equation}{section}

\begin{titlepage}

	\maketitle

	\thispagestyle{empty}

	\vfill

	\abstract{ \normalfont{}\noindent{}%
          The utility of the non-relativistic large-charge \ac{eft} for
          physical systems, and neutron matter in particular, relies on controlled
          Schr\"odinger-symmetry breaking deformations due to
          scattering length and effective-range effects in the
          two-body system. A recently-found exact solution of the
          large-charge system is used to compute these effects for
          two-point correlation functions of large-charge operators in
          perturbation theory around the large-charge ground
          state. Notably, the leading effective-range effects are
          found to enter at second order in the effective range, in
          agreement with analogous calculations in the three-body
          system. The Schr\"odinger-symmetry breaking deformations are
          used --together with input from Quantum Monte Carlo
          simulations-- to address the range of validity of the
          \ac{eft} with deformations both in general and in the
          special case of neutron matter. In particular, it is found
          that nuclear reactions with up to six low-energy neutrons in
          the final state can be described by the large-charge
          \ac{eft} with Schr\"odinger-symmetry breaking.}

\end{titlepage}

\setstretch{1.1}
\tableofcontents

\acused{lo}

\section{Introduction}
\label{sec:intro}

Recent
work~\cite{Hammer:2021zxb,Schaefer:2021swu,Braaten:2023acw,Chowdhury:2023ahp}
has demonstrated that certain nuclear reactions with low-energy
neutrons in the final state can, in principle, be described by an
\ac{eft} whose leading order is a \ac{nrcft} which describes fermions
at unitarity. As scale-invariant matter has no particle
interpretation, a field in a \ac{nrcft} which describes nuclear matter
may be referred to as an ``unnucleus'', a non-relativistic analog
of the ``unparticles'' considered in Ref.~\cite{Georgi:2007ek}.  As
nuclear systems are not Schr\"odinger invariant, in order to establish
the utility of this \ac{eft}, it is essential to consider deformations
about the unitary fixed point.  The s-wave neutron-neutron scattering
length is large, and therefore the consequent Schr\"odinger-symmetry
breaking effects should be small at sufficiently small densities
and/or momentum transfers. These effects have been computed in
Ref.~\cite{Chowdhury:2023ahp} for systems with up to three neutrons in
the final state, and it was found that they are indeed small as
expected. On the other hand, the neutron-neutron effective range is of
natural size, or larger, and therefore one expects that these effects
will constitute the dominant symmetry-breaking contribution. For
reasons that remain mysterious, the nominally-leading effective-range
contributions in three-neutron systems are found to vanish in
conformal perturbation theory~\cite{Chowdhury:2023ahp}.
Quantification of the effective-range corrections thus requires a more
intricate perturbative calculation that has not yet been done in the
three-body sector.

Given the inherent complexity of the few-body wave functions, in
systems with many neutrons in the final state, it may prove more
efficient to work in an \ac{eft} which describes the superfluid state
of neutron matter~\cite{Greiter:1989qb,Son:2005rv}. Recent work has
shown that correlation functions in this \ac{eft} can be computed
systematically in a large-charge
expansion~\cite{Favrod:2018xov,Kravec:2019djc,Hellerman:2020eff,Pellizzani:2021hzx,Hellerman:2021qzz,Hellerman:2023myh},
where the conformal dimension of operators is efficiently computed
using the state-operator correspondence~\cite{Nishida:2007pj}.
Unfortunately, it is not immediately clear how to adapt the
state-operator correspondence to account for Schr\"odinger-symmetry
breaking deformations. This renders it challenging to compute the
symmetry-breaking contributions to the correlation functions.
One of the aims of this work is to show how this can be done systematically.%
\footnote{The near-conformal dynamics due to a small dilaton mass in a linear
realization of the large-charge \ac{eft} has been explored in both the
relativistic~\cite{Orlando:2019skh} and
non-relativistic~\cite{Orlando:2020idm} cases.}

The large-charge expansion~\cite{Hellerman:2015nra,Gaume:2020bmp} is a powerful tool
to analytically access strongly-coupled \acp{cft} in a sector of large global charge.
In very recent work~\cite{Beane:2024kld}, the large-charge 
master field has been found which allows one to systematically compute
all Schr\"odinger-invariant n-point correlation functions with a
single large-charge insertion in the large-charge limit. This solution
further allows the simple computation of Schr\"odinger symmetry
breaking corrections in the large-charge \ac{eft} due to finite
scattering-length effects in the fundamental theory of fermions near
unitarity~\protect\cite{Beane:2024kld}. It is reassuring that at
charge $Q=3$, these results are found to be consistent with the
results of Ref.~\cite{Chowdhury:2023ahp} which are computed directly
in conformal perturbation theory with the three-body
wavefunctions. 

This work will consider deformations of Schr\"odinger
symmetry due to effective-range effects, and in addition will consider
the general form of the scattering-length and effective-range
deformations in a perturbation theory around the large-charge ground state. 
Perhaps not surprisingly, the
nominally-leading effective-range corrections in the large-charge
\ac{eft} are found to vanish, consistent with the $Q=3$
result. Therefore, second-order effects in the effective range are
computed in perturbation theory.

It may appear that solving the \ac{eom} order-by-order in perturbation
theory around the Schr\"odinger-invariant point is a daunting task;
time-translation invariance is broken by the operator insertions, and
Schr\"odinger invariance is explicitly broken by the finite-range
corrections.  However, the task becomes manageable if a coordinate
transformation is applied that corresponds to a frame change in the
associated \ac{nrcft}, \emph{i.e.} in the undeformed system.  For a
Schr\"odinger-invariant system this transformation shifts the
insertions to times \(\tilde{\tau} = \pm \infty\), preserving the form
of the Lagrange density but introducing a background harmonic
potential (this is the basis of the non-relativistic state-operator
correspondence).  In the case at hand, moreover, the
Schrödinger-breaking couplings become time-dependent. The resulting
system in this \emph{oscillator frame} is non-autonomous, but the
\ac{eom} simplify, allowing one to find explicit closed-form
solutions, which are then transformed back to the initial \emph{flat
  frame}.  In this way one finds an explicit expression for the
correlation function, that in energy-momentum space reads
\begin{equation}
\Im G(E,{\mathbf 0}) = C_0\;E^{\Delta_Q-5/2} \bqty*{ 1  \ +\ {\cal C_Q}\,\pqty*{a\,\sqrt{ME}}^{-1} \ +\ {\cal C''_Q}\,{r_0^2\,{ME}} }\; ,
\label{eq:swscbe8c2}
\end{equation}
where \(\Delta_Q\) is the conformal dimension of the lowest operator of charge \(Q\), and \({\cal C_Q}\) and \({\cal C''_Q}\) are known functions of the charge and of the Lagrange-density parameters.
While \({\cal C_Q}\) has been computed in Ref.~\cite{Beane:2024kld}, a primary goal in what follows is to compute \({\cal C''_Q}\).

\bigskip 

The large-charge \ac{eft} in the Schr\"odinger limit is a \acl{nrcft},
and therefore, a priori, can be applied only to systems at criticality.
However, there is a wide range of physical systems that may be
profitably studied using the deformed theory, as the \ac{eft}
describes all underlying non-relativistic many-body systems that
experience superfluidity (\emph{i.e.} the spontaneous breaking of the
particle number symmetry) and that are by some measure near
criticality.  This is, for example, the case with superfluid atomic
gases.  Using Feshbach resonances, one can tune the underlying
contact forces arbitrarily close to unitarity, and have naturally
vanishingly-small effective-range effects and controllable three-body
effects. A second example, which is not considered in this paper, is a gas of anyons in $2+1$ dimensions which
is deformed away from the Schr\"odinger limit~\cite{Bergman:1993kq}.

Of special relevance here are few--body systems of
neutrons. Such systems have long been of interest to theorists due to
the intriguing possibility of the existence of pure-neutron
nuclei\footnote{See, for instance,
  Refs.~\cite{Maris:2013rgq,Marques:2021mqf}.}. Interest has ramped up
recently due to new experimental efforts, particularly at the
Radioactive Ion Beam Factory at \textsc{riken}, using the \ac{samu}~\cite{Yang:2019zgd,Nakamura}. In particular, \ac{samu} is able
to resolve the energies of the final state neutrons with unparalleled
precision. To date reactions with four neutrons in the final state
have been studied~\cite{Duer:2022ehf}, and systems with six and more
final-state neutrons are being explored~\cite{Nakamura}. The deformed
large-charge \ac{eft} may provide a valuable tool for computing correlation
functions of these systems of low-energy neutrons in the final state
of nuclear reactions; i.e. unnuclear matter~\cite{Hammer:2021zxb,Schaefer:2021swu,Braaten:2023acw,Chowdhury:2023ahp}.

\medskip
It is the primary goal of this paper to provide a quantitative answer
to the question: how close to unitarity must a system of $Q$ fermions
be in order to be profitably described using the deformed large-charge
\ac{eft}? A corollary of this question relevant to neutron systems is:
what is the range of values of $Q$ for which the large-charge \ac{eft}
provides a systematic, controllable, approximation scheme for
describing neutron correlation functions? From the perspective of the
fundamental theory of fermions near unitarity, the existence of the
large-charge \ac{eft} relies on the existence of a well-defined Fermi
surface, which in turn implicitly assumes a gas of fermions in the
thermodynamic limit. Thus, given the many-body nature of the large-charge
\ac{eft}, one should expect that the utility of this \ac{eft} in describing
few-body systems of neutrons will depend critically on the detailed
numerical behavior of the perturbative expansion.
In any event, answering these questions
clearly requires a systematic calculation of the leading
symmetry-breaking effects due to scattering length and effective-range
effects.

\bigskip
This paper is organized as follows. In Section~\ref{sec:fnu}, the
underlying \ac{eft} of fermions near unitarity is reviewed, and the
\ac{nrcft} is defined. Section~\ref{sec:seftgc} in turn reviews the
superfluid \ac{eft} which describes fermions near unitarity in the far
infrared where the sole active \ac{dof} is the Goldstone
boson of spontaneously broken particle-number symmetry. The
symmetry-breaking operators in this \ac{eft} are written down, and,
via a calculation of the energy density of the system, the
coefficients of these operators are shown to be determined by quantum
\ac{mc} simulations of the near-unitary Fermi gas. The
large-charge \ac{eft} in sectors of fixed charge $Q$ is then reviewed
in Section~\ref{sec:cfalc}. Of special importance is the recently
discovered master-field solution, which serves as the
Schr\"odinger-invariant Goldstone boson field about which 
perturbation theory is constructed.
Sections~\ref{sec:first-order-range} and~\ref{sec:nnlo-range} develop
perturbation theory to first- and second-order in the
effective range, respectively. In computing the large-charge
correlation functions, even in the symmetry limit, special attention
must be paid to divergences that appear at the boundary of Euclidean
time. Section~\ref{sec:regularization2pt} considers various
regularization schemes in order to properly separate regularization
artifacts from physics.
While the leading scattering-length corrections were obtained in Ref.~\cite{Beane:2024kld},
Section~\ref{sec:losls} recovers these corrections using the new methodology.
The final technical steps, continuation back
to Minkowski space and Fourier transformation of the correlation
functions to energy-momentum space are described in
Section~\ref{sec:mscf}.  Finally, in Section~\ref{sec:um} an analysis
of the validity of perturbation theory is carried out generally and in
the special case of neutron matter. Section~\ref{sec:conc} summarizes
and concludes.  In order to relieve the narrative of clutter, some
essential material has been organized into a series of appendices.

\section{Fermions near unitarity: EFT definition}
\label{sec:fnu}

\begin{figure}[!ht]
\centering
\includegraphics[width = 0.9\textwidth]{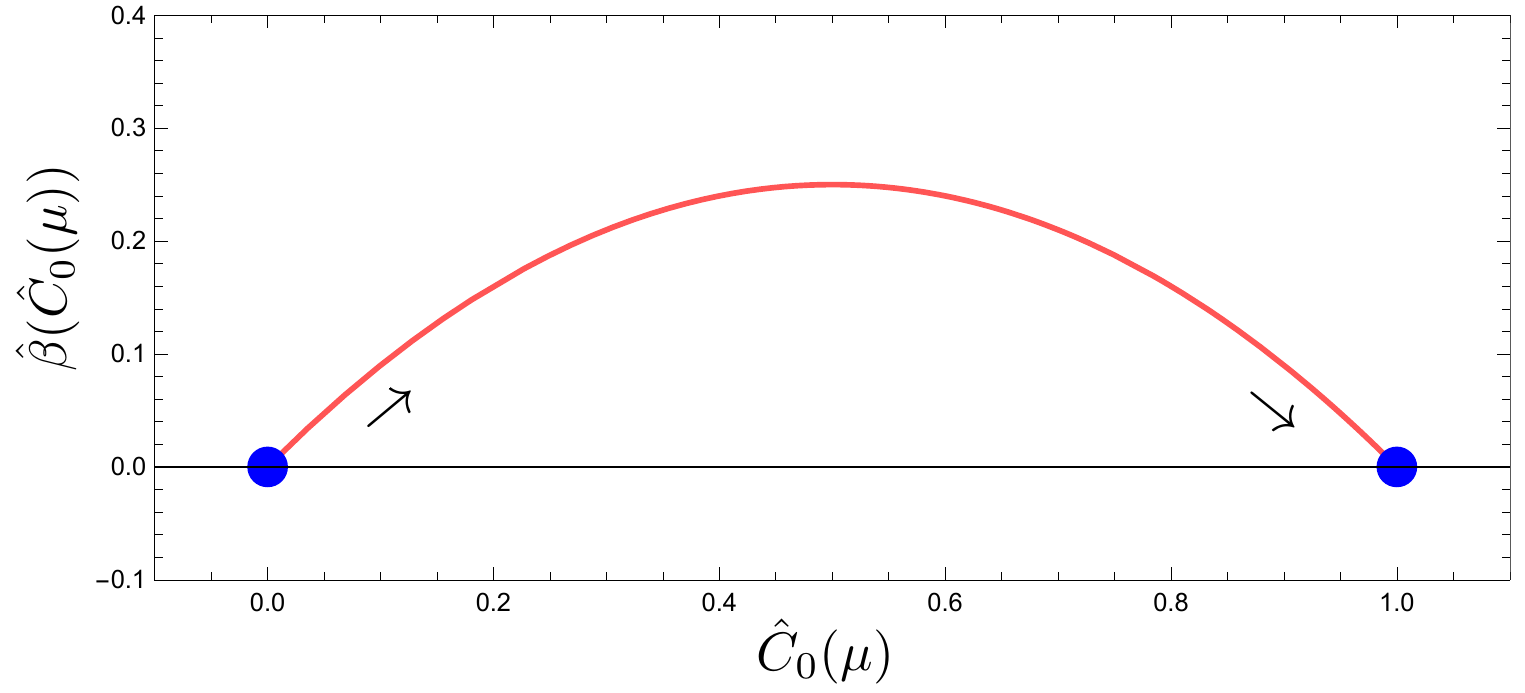}
\caption{The beta-function of Eq.~(\ref{eq:c0betafn}) plotted as a function of ${\hat C}_0(\mu)$. The blue dots are the \ac{rg} fixed points.
  The arrows indicate the direction of increasing $\mu$; the beta-function curve evolves from $\mu=0$ (the trivial fixed point) to $\mu=\infty$ (the unitary fixed point).}
  \label{fig:eftc0beta}
\end{figure}

Consider a system of spin-$1/2$ fermions which interact via two-body
contact forces.  At very low energies, where derivative interactions
can be ignored, the Lagrange density can be expressed as
\begin{eqnarray}
  {\cal L}  &=&
       \psi_\sigma^\dagger \bigg\lbrack i\partial_t + \frac{\nab^{\,2}}{2M}\bigg\rbrack \psi_\sigma +  \frac{1}{C_0} s^\dagger s + \psi_\downarrow^\dagger \psi_\uparrow^\dagger  s + s^\dagger \psi_\uparrow \psi_\downarrow\ ,
  \label{eq:lagdimeron}
\end{eqnarray}
where the field $\psi_\sigma^\dagger$ creates a fermion of spin $\sigma=\uparrow,\downarrow$, $s$ is an auxiliary field, and $C_0$ is a bare low-energy constant.
In what follows these fermions will be taken to be neutrons. Consider neutron-neutron scattering. Below inelastic
thresholds, the s-wave phase shift is given by the effective-range expansion
\begin{equation}
k \cot\delta(k)\ =\  -\frac{1}{a} \ +\  \frac{1}{2} r_0 k^2 \ +\  v_2  k^4 \ +\  \order{k^6} \ ,
\label{eq:eregen}
\end{equation}
where $k=\sqrt{ME}$ is the on-shell center-of-mass momentum,
$a$ is the scattering length, $r_0$ is the effective range, and $v_2$ is a shape parameter. 
In dimensional regularization with the power-divergence subtraction scheme~\cite{Kaplan:1998tg,Kaplan:1998we} and
renormalized at the scale $\mu$, the relation between the low-energy
constant $C_0$ and the scattering length is given by
\begin{equation}
  C_0(\mu) \ =\ \frac{4 \pi}{M} \frac{1}{{1}/{a}-{\mu}} \ .
\label{eq:c0run}
\end{equation}
There is a non-trivial \ac{uv} fixed point at $C_0=C_\star$, corresponding
to a divergent scattering length. There is also a trivial fixed point at $C_0=0$, corresponding to free particles ($a=0$).
Rescaling the couplings to ${\hat C}_0 \equiv C_0/C_\star$, the beta-function for the rescaled coupling is
\begin{equation}
  {\hat\beta({\hat C}_0)}\ =\  \mu \frac{d}{d\mu} {\hat C}_0(\mu) \ =\ -{\hat C}_0(\mu)\left({\hat C}_0(\mu)-1\right) \ ,
   \label{eq:c0betafn}
\end{equation}
which has fixed points at ${\hat C}_0=0$ and $1$, as shown in Fig.~\ref{fig:eftc0beta}. The coupling is near the
trivial fixed point for $\mu < 1/|a|$, and near the non-trivial fixed point for $\mu
> 1/ |a|$.

Expanding about the \ac{uv} fixed point, the system of low-energy fermions is described by a
\ac{nrcft}, defined by the Lagrange
density
\begin{eqnarray}
  {\cal L}_{CFT}  &=&
       \psi_\sigma^\dagger \bigg\lbrack i\partial_t + \frac{\nab^{\,2}}{2M}\bigg\rbrack \psi_\sigma +  \frac{1}{C_\star} s^\dagger s + \psi_\downarrow^\dagger \psi_\uparrow^\dagger  s + s^\dagger \psi_\uparrow \psi_\downarrow\ .
       \label{eq:lagdimeroncft}
\end{eqnarray}
Consider now the inclusion of small Schr\"odinger-breaking effects in the fundamental theory.  Using Eq.~(\ref{eq:lagdimeron}) and Eq.~(\ref{eq:c0run}) one can write
\begin{eqnarray}
  {\cal L}  &=&   {\cal L}_{CFT} \ +\ \frac{M}{4\pi a} s^\dagger s \ -\ \frac{M^2 r_0}{8\pi} s^\dagger\left( i\overleftrightarrow{\partial_t} + \frac{\nab^{\,2}+\nabl^{\,2}}{4M} \right) s \ ,
  \label{eq:lagdimerondecomp}
\end{eqnarray}
where, in addition, effective-range corrections have been
included~\cite{Beane:2000fi,Chowdhury:2023ahp}. Note that the field
$\psi_\sigma$ has conformal dimension $3/2$ while the field $s$ at
unitarity has conformal dimension $2$~\cite{Nishida:2007pj}. The scattering length
corrections therefore enter via a relevant dimension-$4$ operator, and
the effective-range corrections enter via an irrelevant dimension-$6$
operator. Therefore, the operators are formally scale invariant if
$a^{-1}$ and $r$ are assigned conformal dimensions $1$ and $-1$, respectively.

\section{Superfluid EFT}
\label{sec:seftgc}

\subsection{Basics of the Euclidean EFT}

Many neutrons at and near unitarity are superfluid and therefore are described in the \ac{ir} by an \ac{eft} of the Goldstone boson, $\theta(x)$,
of spontaneously broken particle number~\cite{Greiter:1989qb,Son:2005rv}. In Euclidean space, the partition function is
\begin{eqnarray}
Z \ =\  \int \DD{\theta} \exp\left(-S \right) \ =\ \int {\cal D}\theta \exp\left(-\int \dd{\tau} \dd^3{\mathbf x} {\cal L} \right) \ ,
\label{eq:efoeft1}
\end{eqnarray}
where, at \ac{lo}, 
\begin{eqnarray}
  {\cal L}_{LO} \ =\ -c_0\,M^{3/2} X^{5/2}  \ ,
\label{eq:efoeft3}
\end{eqnarray}
with
\begin{eqnarray}
X \ =\ i \partial_\tau \theta \, -\,\frac{(\partial_i\theta)^2}{2M} \, .
\label{eq:efoeft4}
\end{eqnarray}
One readily checks that ${\cal L}_{LO}$ is Schr\"odinger invariant and therefore defines
a \ac{nrcft}~\cite{Greiter:1989qb,Son:2005rv}. 
The density and Hamiltonian density are given by, respectively,
\begin{eqnarray}
	\rho & =&  -\frac{\delta {\cal L}}{\delta X} \  \ \ \ ,\  \ \ \  {\cal H} \ =\ {\cal L} \; -\; \dot\theta \frac{\partial{\cal L}}{\partial\dot\theta} \ ,
\label{eq:efoeft7}
\end{eqnarray}
where $\dot\theta\equiv \partial_\tau\theta$. The Euler--Lagrange equations take the form
\begin{eqnarray}
    \partial_\tau \rho \, +\,   \frac{1}{M} \partial_i \left(i {\partial_i\theta}\rho \right) \,=\, 0 \; .
\end{eqnarray}
At \ac{lo}, one finds the  solution\footnote{From this equation forward, $\mu$ denotes the chemical potential.}
\begin{equation}
	\theta  \;=\; -i\,\mu\,\tau \ \ \  , \ \ \  X   \;=\; \ \mu\, \ ,
\label{eq:efoeft6}
\end{equation}
which describes the homogeneous ground state of the system.

\subsection{Symmetry breaking operators by matching}

As the Goldstone field $X$ carries conformal dimension $2$, a spurion analysis gives the leading symmetry-breaking operators in the superfluid \ac{eft},
\begin{eqnarray}
{\cal L}_{\textsc{sb}} \; =\;  -g_1\,a^{-1}{M X^2} \; -\; g_2\,a^{-2}M^{1/2} X^{3/2} \; -\; h_1\,r_0 {M^2 X^3} \; -\; h_2\,r_0^2 {M^{5/2} X^{7/2}}\; +\; \ldots \; ,
  \label{eq:Lsbfunit}
\end{eqnarray}
where the $g_i$ and the $h_i$ are dimensionless constants.%
\footnote{Note that this is the Euclidean Lagrange density, while the constants
  have been defined with a plus sign in Minkowski space. The
  parameters $g_{1,2}$, defined in Ref.~\cite{Beane:2024kld}, differ
  by an overall sign.}  Here it is assumed that the scattering length
and effective range effects are tuned independently; i.e., no mixed
operators are considered. The ordering of these operators within the large-charge \ac{eft} will
be considered in detail below. Note that shape parameter corrections
enter at $\order{r_0^3}$ and therefore will not be considered in this work. In the homogeneous ground state, with
$M=1$, the grand-canonical potential is read off from the total Lagrange
density as:
\begin{eqnarray}
\Omega(\mu) = -c_0 \mu^{5/2} - g_1 a^{-1} \mu^2  - g_2 a^{-2} \mu^{3/2} - h_1 r_0 \mu^{3} - h_2 r_0^2 \mu^{7/2} + \ldots .
  \label{eq:GCpot}
\end{eqnarray}
The constants $g_i$ and the $h_i$ are numbers of order unity which determine quantitatively the
leading effects of the deformation of Schr\"odinger symmetry due to finite $a$ and non-vanishing $r_0$. 
As will be shown in the next section, these numbers have been
determined using quantum \ac{mc} simulations.

\subsection{Energy per particle}

\begin{figure}[!ht]
\centering
\includegraphics[width = 0.8\textwidth]{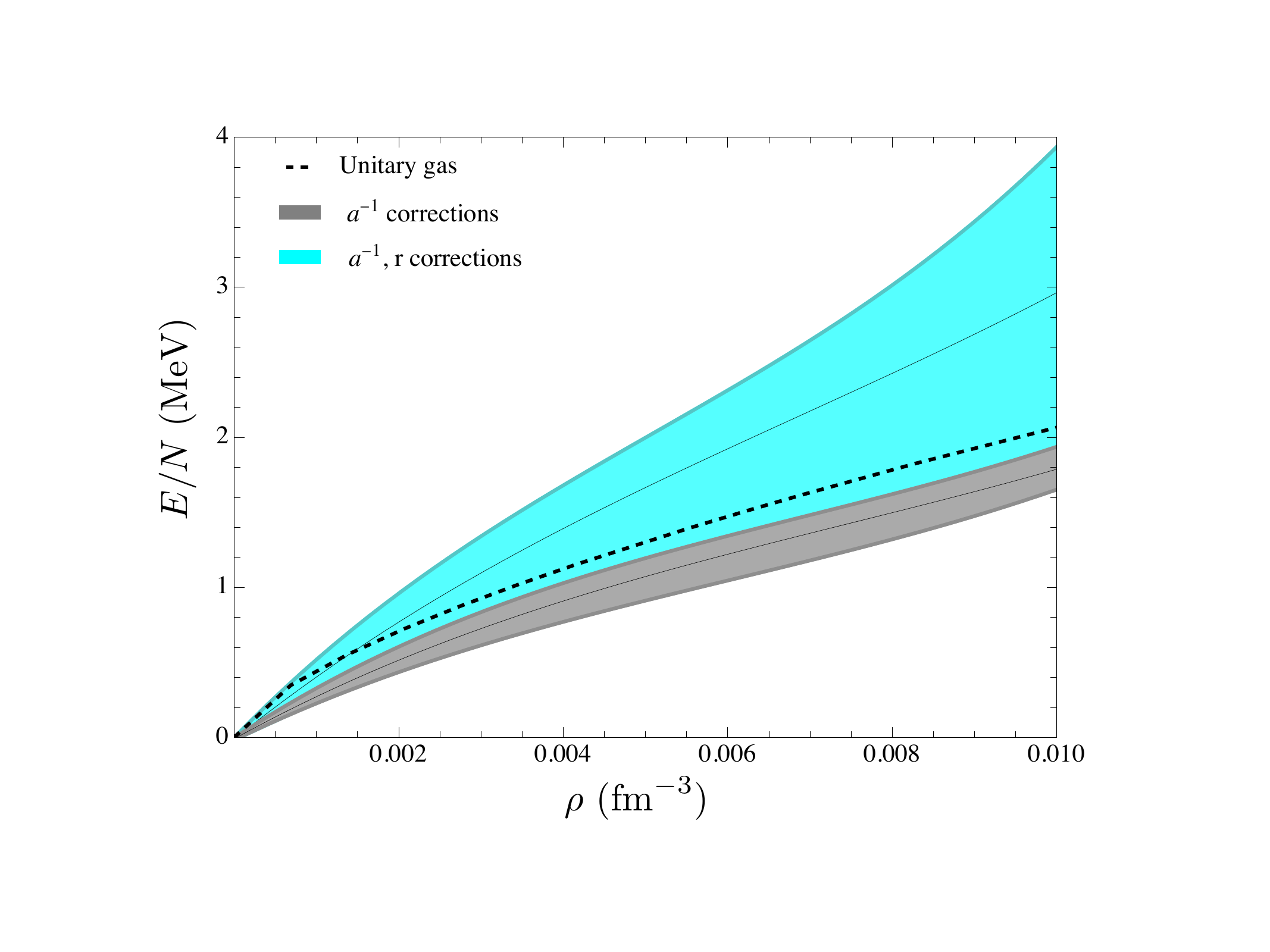}
\caption{Energy per particle in neutron matter at low densities, as taken from Eq.~(\ref{eq:eoNunitarity}). The error bands are from propagation of the quantum \ac{mc} uncertainties.}
  \label{fig:lowrho}
\end{figure}

Using dimensional analysis to write down deformations away from
unitarity, the energy-per-particle $E/N$ of the interacting Fermi gas
in the near-Schr\"odinger limit can be written at very-low densities
as~\cite{PhysRevA.84.061602,Bulgac:2005zza,vanKolck:2017jon}
\begin{equation}
E/N \ =\  \frac{3}{5} \frac{\kf^2}{2M}  \left( \xi \ -\  \frac{\zeta}{\kf a} \ -\  \frac{\zeta_2}{\kf^2 a^2}\ +\ \ldots \ +\ \eta\, \kf r_0 \ +\ \eta_2\, \kf^2 r_0^2 \ + \ \ldots    \right) \ .
\label{eq:eoNunitarity}
       \end{equation}
Here the various dimensionless universal parameters have been
determined using quantum \ac{mc} simulations. From
Ref.~\cite{PhysRevA.84.061602}, $\xi = 0.372(5)$ (Bertsch parameter)
and $\eta=0.12(3)$. The quadratic range corrections have been
studied in Ref.~\cite{Forbes:2012ku}; the average of two determinations
gives $\eta_2=-0.03(2)$.
From the simulation data in Ref.~\cite{Chang:2004zza}, it is
straightforward to extract $\zeta =-0.68(15)$ and $\zeta_2=3.6(10)$. For the case of neutron matter, at densities corresponding to interparticle separations greater
than the pion Compton wavelength, the energy-per-particle with the quantum \ac{mc} determination of the deformation away from unitarity is shown in Fig.~\ref{fig:lowrho}.

It is straightforward to obtain Eq.~(\ref{eq:eoNunitarity}) in the superfluid \ac{eft}. The energy density can be obtained from the
grand-canonical potential by the (Euclidean) Legendre transform,
\begin{equation}
\Omega(\mu) \ = \ \edens(\mu) \ +\ \mu \rho \ .
\label{legtrans}
\end{equation}
Note that while the density $\rho$ is fixed in terms of $\kf$, the chemical potential $\mu$ is shifted away from the Fermi energy by the symmetry-breaking effects. That is,
\begin{equation}
\rho \ = \ \frac{\kf^3}{3\pi^2} \ =\ -\frac{\delta \Omega(\mu)}{\delta \mu} \ .
\label{fermid}
\end{equation}
One then finds Eq.~(\ref{eq:eoNunitarity}) with
\begin{align}
c_0  &= \frac{2^{5/2}}{15\pi^2 \xi^{3/2}} = 0.168(3) \ ,  \\
g_1  &= \frac{2 \zeta}{5\pi^2 \xi^{2}}= -0.20(4)\ \ , & g_2  &= \frac{\sqrt{2}}{25\pi^2 \xi^{5/2}}\left(4\zeta^2+5\zeta_2 \xi \right)= 0.58(6) \ ,\\
  h_1   &= \frac{4 \eta}{5\pi^2 \xi^{3}} = 0.19(5) \ ,  &  h_2  &= \frac{4\sqrt{2}}{25\pi^2 \xi^{9/2}}\left(9\eta^2-5\eta_2 \xi \right)= 0.38(15)  \ .
                                                           \label{eq:g0g1h1}
\end{align}                                                           
These values of the universal constants will be used below in the quantitative determination of the deformed correlation functions.

\section{Two-point function at large charge}
\label{sec:cfalc}

This section briefly reviews the main developments of
Ref.~\cite{Beane:2024kld} as this provides the basis of the
perturbation theory that follows. In the large-charge \ac{eft}, the charge-$Q$ primary operator of
conformal dimension $\Delta$ is
\begin{equation}
{\cal O}_{\Delta,Q} = {\cal N} X^{\Delta/2} \exp\left(i Q \theta  \right)\, ,
\label{eq:swseft2}
\end{equation}   
where ${\cal N}$ is a normalization constant. The two-point function which evolves superfluid matter of charge $Q$ and \textit{ a priori} unknown
conformal dimension $\Delta$ from source point $x_2$ to source point $x_1$ in the Schr\"odinger limit, is defined via the path integral
\begin{equation}
G_Q(x_1;x_2) \ =\ G_Q(\tau_1,\mathbf{x}_1;\tau_2, \mathbf{x}_2)  \ =\  \int \DD{\theta} {\cal O}_{\Delta,Q}(x_2) {{\cal O}}_{\Delta,-Q} (x_1)\, e^{-\int \dd^4{x} {\cal L}_{LO}} \ .
\label{eq:swseft3}
\end{equation}   
Schr\"odinger symmetry completely constrains the two-point function to be of the form~\cite{Hagen:1972pd,Niederer:1972zz,Henkel:1993sg}
\begin{equation}
    G_Q(x_1,x_2) =  \mathcal{N}^{2} {\tau_{12}^{-\Delta}} \exp\pqty*{-\frac{Q M \mathbf{x}_{12}^2}{2 \tau_{12}}} \ ,
\end{equation}
where $\tau_{12}\equiv \tau_1-\tau_2$, $\mathbf{x}_{12}\equiv \mathbf{x}_1-\mathbf{x}_2$.
In the presence of sources, the \ac{eom} acquires a source term so that
\begin{eqnarray}
    \partial_\tau \rho \, +\,   \frac{1}{M} \partial_i \left(i {\partial_i\theta}\rho \right) \,=\,
Q\,\big\lbrack\delta^4(x-x_2)-\delta^4(x-x_1)\big\rbrack \,.
\end{eqnarray}
The master-field solution to the \ac{eom} and the saddle point location in the presence of sources is given by~\cite{Beane:2024kld,Son:2021kkx}
\begin{eqnarray}
  \theta_s(\tau, \mathbf{x} | \tau_1,\mathbf{x}_1;\tau_2, \mathbf{x}_2) &=&  \frac{i}{2} \gamma \log\left(\frac{\tau_1-\tau}{\tau-\tau_2}\right) \;-\; \frac{i}{4}
  \Bigg\lbrack  \frac{(\mathbf{x}-\mathbf{x}_2)^2}{(\tau-\tau_2)}\ -\ \frac{(\mathbf{x}-\mathbf{x}_1)^2}{(\tau_1-\tau)}   \Bigg\rbrack \ ,
\label{eq:swseft16He}
\end{eqnarray}
where the anomalous dimension $\gamma$ is
\begin{eqnarray}
\gamma &=&  \frac{\mu}{2}\tau_{12} \ =\ {3^{1/3}\xi^{1/2}}Q^{1/3} \,.
\label{eq:swseft27}
\end{eqnarray}   
Now the two-point function evaluated at the master-field solution is
\begin{equation}
  \label{eq:two-point-function}
    G_Q(x_1,x_2) = \lim_{\epsilon \to 0} \eval*{\Op{\Delta}{Q}(\tau_2 + \epsilon, \mathbf{x}_2) \Op{\Delta}{-Q} (\tau_1 - \epsilon, \mathbf{x}_1)  \exp\left[-\int_{\tau_2 + \epsilon}^{\tau_1-\epsilon} \dd{\tau} \int \dd^3{\mathbf{x}} {\cal L}_{LO}[\theta]\right] }_{\theta = \theta_s} ,
\end{equation}
where the temporal boundaries have been shifted  by $\epsilon$.
The existence of the $\epsilon\to 0$ limit provides a constraint that is used to compute the conformal dimension \(\Delta\)~\cite{Beane:2024kld}.

One finds at the saddle,
\begin{equation}
  \label{eq:saddle-action}
  \begin{aligned}
    S_{\saddle}[\theta_s] &= - \frac{3^{1/3} \xi^{1/2}}{4} Q^{4/3} \log\left( \frac{\tau_{12}}{\epsilon} \right) \ .
  \end{aligned}
\end{equation}
Substituting the saddle solution into the expression for the operators one then finds
\begin{equation}
  G_Q(x_1;x_2)\sim  \epsilon^{Q \gamma - \Delta - \frac{3^{1/3} \xi^{1/2}}{4} Q^{4/3}}  {\tau_{12}^{-Q \gamma  +\frac{3^{1/3} \xi^{1/2}}{4} Q^{4/3} }} e^{-\frac{Q M \mathbf{x}_{12}^2}{2 \tau_{12}}} \ .
\end{equation}
Finally, absence of the divergence for \(\epsilon \to 0\) determines the conformal dimension
\begin{equation}
    \Delta_Q = Q \gamma - \frac{3^{1/3} \xi^{1/2}}{4} Q^{4/3} = \frac{3^{4/3}}{4} \xi^{1/2} Q^{4/3}  \ ,
  \label{eq:cdLO}
\end{equation} 
in agreement with the state-operator correspondence~\cite{Kravec:2019djc}.

The goal of this paper is to perturbatively compute the leading symmetry-breaking corrections to this correlation function, with 
\begin{equation}
 {\cal L}[\theta] \ =\  {\cal L}_{LO}[\theta] \ +\ {\cal L}_{\textsc{sb}}[\theta] \ .
\end{equation}
Naively, the leading symmetry-breaking corrections are evaluated by computing the symmetry-breaking action at the saddle-point solution. 
For instance, for the two-point function, one expects
\begin{equation}
G(x_1;x_2) \ = \  G_Q(x_1;x_2) e^{-S_{\textsc{sb}}[\theta_s]}\ .
\label{eq:cpt1}
\end{equation}   
While in practice this gives the correct result for the scattering-length corrections~\cite{Beane:2024kld}, in general this procedure fails, as there are non-vanishing boundary terms
in the effective action which must be included. A simple example of an action whose perturbative expansion contains boundary terms is provided in Appendix~\ref{sec:boundaryTerm}.

\section{First-order range correction}%
\label{sec:first-order-range}

\subsection{Perturbative expansion defined}
\label{sec:ped}

This section aims to describe the effect of a finite effective interaction range \(r_0\) on the form of the two-point function.
The approach is perturbative, and one begins with the Lagrange density
\begin{equation}\label{eq:L-breaking}
  {\cal L} = -c_0 X^{5/2}  - h_1 r_0 X^3  - h_2 r_0^{2} X^{7/2} + \cdots .
\end{equation}
First, the parametric range of validity of the approximation should be
established.  At the saddle solution, Eq.~(\ref{eq:swseft16He}), corresponding to the insertion of two
operators at distance \(\tau_{12}\) in the temporal direction, the
operator \(X_{\saddle}\equiv X[\theta_s]\) scales as \(X_{\saddle} =
\order{\mu} = \order{\gamma/\tau_{12}}\).  Thus, the \(r_0\) term can
be treated perturbatively as long as
\begin{equation}
  r_0 \sqrt{\mu} = r_0 \sqrt{\frac{\gamma}{\tau_{12}}} \ll 1 .
\end{equation}
A hierarchy with respect to Schr\"odinger-invariant corrections to Eq.~(\ref{eq:L-breaking}) should also be established. One may assume, for instance, that the leading
effective-range corrections are much larger than the first subleading correction in the large-charge expansion of the pure Schrödinger system.
It is known~\cite{Son:2005rv,Kravec:2018qnu} that this latter correction is suppressed by \(1/\gamma^2\) with respect to the \(c_0\) term.
Therefore, in addition one has
\begin{equation}
    r_0 \sqrt{\mu} = r_0 \sqrt{\frac{\gamma}{\tau_{12}}} \gg \frac{1}{\gamma^2}.
\end{equation}
Using the fact that \(\gamma = \order{Q^{1/3}}\), it then follows that the effective range in units of \(\tau_{12}\) should satisfy
\begin{equation}
  \frac{1}{Q^{5/6}} \ll \frac{r_0}{\sqrt{\tau_{12}}} \ll \frac{1}{Q^{1/6}}.  
\end{equation}
It will be seen below that the first non-vanishing correction due to the effective range is of order \(r_0^2\), and, therefore, by the same argument one has
\begin{equation}
  \frac{1}{Q} \ll \frac{r_0^2}{\tau_{12}} \ll \frac{1}{Q^{1/3}} . 
\end{equation}
Hence, formally, the perturbative expansion developed below treats the limit \(r_0 \to 0\), \(\mu \to \infty\) with the product \(r_0 \sqrt{\mu} \equiv \kappa\) held fixed and small.

An analogous analysis applies in consideration of the perturbation theory with a finite scattering length \(a\).
In this case, a consistent hierarchy is
\begin{equation}
  \frac{1}{Q^{1/2}}  \ll \frac{\sqrt{\tau_{12}}}{a} \ll Q^{1/6},
\end{equation}
where formally, the perturbative expansion treats the limit \(a \to \infty\), \(\mu \to \infty\) with the product \(a^{-1} \sqrt{\mu}\) held fixed and small.
The leading symmetry-breaking corrections due to a finite scattering length were obtained in Ref.~\cite{Beane:2024kld},
and a generalization of this solution will be discussed below.

\subsection{EOM in the oscillator frame}
\label{sec:eom-oscillator-frame}

The goal is to perturbatively compute the two-point function with Schr\"odinger-symmetry breaking;  \emph{i.e.}
\begin{equation}
G_Q(x_1;x_2) \ =\  \int \DD{\theta} {\cal O}_{\Delta,Q}(x_2) {{\cal O}}_{\Delta,-Q} (x_1)\, e^{-\int \dd^4{x} {\cal L}} \ ,
\label{eq:swseft3b}
\end{equation}   
with the Lagrange density given in Eq.~(\ref{eq:L-breaking}). While
this Lagrange density breaks scale and conformal invariance, it
preserves Galilean invariance (including translations and rotations) since
the symmetry-breaking operators commute with Galilean boosts. It
follows that the two-point function of primary operators must take the
form~\cite{Chowdhury:2023ahp,Henkel:1993sg}
\begin{equation}
	G_Q(x_1;x_2) = f(\tau_{12}) \exp\pqty*{-\frac{Q\mathbf{x}_{12}^2}{2\tau_{12}}},
\end{equation}
where $f$ is an arbitrary function. Hence, the dependence on the
spatial component of the insertions is completely fixed in the deformed theory. Because of
this, in the following one can specialize to the simpler case where operator insertions are taken at
$\mathbf{x}_1=\mathbf{x}_2=0$ and $\tau_1=1/\omega$, $\tau_2=-1/\omega$.

The strategy is then to compute the semiclassical field configuration
$\theta(\tau, \mathbf{x} | -\sfrac{1}{\omega},0 ;\sfrac{1}{\omega}, 0)$
resulting from the insertions.  This choice
does not result in any loss of generality.
Since the deformed system retains Galilean invariance, the field configuration
corresponding to generic insertions may be obtained using a
Galilean boost that maps \((-\sfrac{1}{\omega}, 0)\) into \((\tau_2, \mathbf{x}_2)\) and \((\sfrac{1}{\omega}, 0)\) into \((\tau_1, \mathbf{x}_1)\) (see \emph{e.g.}~\cite{Favrod:2018xov} for the transformation properties of \(\theta\)):
\begin{multline}
  \theta(\tau, \mathbf{x} | \tau_2, \mathbf{x}_2 ; \tau_1, \mathbf{x}_1) = \theta(\tau + \tfrac{\tau_1 + \tau_2}{2}, \mathbf{x} + \omega \tau \tfrac{\mathbf{x}_1 - \mathbf{x}_2}{2} + \tfrac{\mathbf{x}_1 + \mathbf{x}_2}{2} | -\sfrac{1}{\omega},0 ; \sfrac{1}{\omega}, 0) + \\
  - \frac{i \omega}{2} \pqty{\mathbf{x}_1 - \mathbf{x}_2} \cdot \pqty*{  \mathbf{x}- \frac{\omega \tau}{4} \pqty{\mathbf{x}_1 - \mathbf{x}_2} }.
\end{multline}

Even with the simplified source locations, the \ac{eom} in the
presence of symmetry breaking is highly complex.  A fruitful idea is
to map the problem to a different frame where the \ac{eom} become
tractable, and then revert to the original frame.  As explained above,
the focus will be on the case \(\mathbf{x}_1 = \mathbf{x}_2 =
0\), and \(\tau_1 = - 1/\omega \) and \(\tau_2 = 1/\omega\).  This choice is
convenient because then one can transform to the oscillator frame, where the
insertions are at \(\tilde \tau = \pm \infty\).  The relevant change of
variables is given in~\cite{Goldberger:2014hca,Beane:2024kld}:
\begin{align}\label{eq:varchange}
  \begin{cases}
    \omega \tau = \tanh(\omega \tilde\tau) \\
    \mathbf{x} = \frac{\tilde{\mathbf{x}}}{\cosh(\omega \tilde \tau)}
  \end{cases} \!\!\!&, &\!\!\!
  \begin{cases}
    \omega \tilde \tau = \arctanh(\omega \tau) \\
    \tilde{\mathbf{x}} = \frac{\mathbf{x}}{\sqrt{1 - \omega^2 \tau^2}}
  \end{cases}
\!\!\!  &,&\!\!\! \theta(\tau, \mathbf{x}) = \tilde \theta(\tilde \tau, \tilde{\mathbf{x}}) - \frac{i}{4} \pqty*{ \frac{x^2}{\tau + 1/\omega} + \frac{x^2}{\tau - 1/\omega}}.
\end{align}
Consider the Lagrange density with the nominally leading effective-range corrections.
The \(X^{5/2}\) term is Schrödinger invariant, while the \(X^3\) term is not, and therefore in the oscillator frame the Lagrange density reads
\begin{equation}
    \tilde {\cal L} = c_0 \tilde X^{5/2} + r_0 h_1 \cosh(\omega \tilde\tau) \tilde X^3 ,
\end{equation}
where now
\begin{equation}
  \tilde X = i \del_{\tilde\tau} \tilde \theta - \frac{\omega^2 \tilde x^2}{2} - \frac{1}{2} (\del_i \tilde\theta)^2 .  
\end{equation}
Note that the symmetry-breaking term has a time-dependent coupling (the Hamiltonian system is non-autonomous).

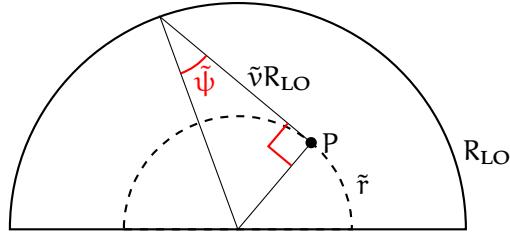
\begin{figure}
  \centering
  \usetikzlibrary{angles, quotes}
  \begin{tikzpicture}
    \draw[thick] (-3,0) -- (3,0) arc(0:180:3) --cycle;
    \draw[dashed,thick] (-1.5,0) -- (1.5,0) arc(0:180:1.5);

    \coordinate (A) at (0,0); %
    \coordinate (B) at (110:3); %
    \coordinate (C) at (50:1.5); %

    \draw[] (C) -- (A) -- (B) -- cycle
      pic ["\(\tilde \psi\)",draw,red,thick,angle radius=0.8cm, angle eccentricity=1.3] {angle}
      pic [draw,red,thick,angle radius=0.4cm] {right angle = A--C--B};

    \node [circle,fill,inner sep=1.5pt] at (C) {};
    \node[right] at (C) {\(P\)};
    \node[] at (20:1.75) {\(\tilde r\)};
    \node[right] at (0,2) {\(\tilde v R_{\ac{lo}}\)};
    \node[right] at (20:3) {\( R_{\ac{lo}}\)};

  \end{tikzpicture}
  
  \caption{The angle \(\tilde \psi\) identifies points \(P\) at
    distance \(\tilde r\) from the center of the droplet (dashed
    circle).  \(\tilde \psi = 0\) is the center of the drop and
    \(\tilde \psi = \pi/2\) is the \ac{lo} edge, where the triangle
    degenerates into a segment.}
  \label{fig:psi-angle}
\end{figure}

The \ac{eom} has the same form as for \(r_0 = 0\), since the Lagrangian depends on \(\tilde\theta\) only via \(\tilde X\):
\begin{equation}
  i \partial_{\tilde\tau} \fdv{\tilde L}{\tilde X} - \del_i \pqty*{\partial_i \tilde\theta \fdv{\tilde L}{\tilde X}} = 0.
\end{equation}
Now consider isotropic solutions, depending only on \(\tilde x = \abs{\tilde{\mathbf{x}}}\), and introduce the variable
\begin{equation}
  \tilde v^2 = 1 - \frac{\omega^2}{2 \mu} \tilde x^2 = 1 - \frac{\tilde x^2}{R_{\ac{lo}}^2},
\end{equation}
where \(R_{\ac{lo}}\) is the size of the droplet at
\ac{lo}~\cite{Son:2005rv,Kravec:2018qnu}, which is spanned by \(0 <
\tilde v < 1\), where \(\tilde v = 1\) is the center and \(\tilde v = 0\) the
border.  In the following it will be convenient to rewrite \(\tilde
v\) in terms of an angular variable \(\tilde v = \cos(\tilde \psi)\).
The distance from the center of the droplet is \(\tilde r
= \abs{\tilde {\mathbf{x}}} = R_{\ac{lo}} \sin(\tilde \psi)\), and it is apparent
that \(\tilde \psi = 0\) is the center, and
\(\tilde \psi = \pi/2\) is the \ac{lo} edge (see
Fig.~\ref{fig:psi-angle}).  As it will be necessary to integrate over the
droplet, it is useful to express the volume element in the various
coordinates:
\begin{equation}
  \dd^3{\tilde x} = \dd{\Omega} \tilde r^2 \dd{\tilde{r}} = - R_{\ac{lo}}^3 \dd{\Omega} \sqrt{1-\tilde v^2} \tilde v \dd{\tilde v} =  R_{\ac{lo}}^3 \dd{\Omega} \sin^2(\tilde \psi) \cos(\tilde \psi) \dd{\tilde{\psi}} ,
\end{equation}
where \(\dd{\Omega}\) is the measure over the two-sphere.

It is convenient to express the \(\order{r_0}\) solution in the form
\begin{equation}
  \tilde \theta(\tilde \tau, \tilde v) = \tilde \theta_0(\tilde \tau, \tilde v) + \frac{\mu^{3/2}}{\omega} r_0 \tilde \theta_1(\tilde \tau, \tilde v) 
\end{equation}
and expand the \ac{eom} to first order in \(\mu^{1/2} r_0 \ll 1\).
The problem has three dimensionful quantities, \(r_0\), \(\mu\) and \(\omega\), from which one obtains two dimensionless ones \(\mu^{1/2} r_0 \) and \(\mu/\omega\).

Recall that the leading-order solution is
\begin{equation}
  \tilde \theta_0 (\tau, \tilde v) = - i \mu \tilde \tau 
\end{equation}
and the \(\order{\mu^{1/2}r_0}\) \ac{eom} takes the simple form
\begin{equation}\label{eq:inho-r}
   \frac{3 \tilde v}{\omega^2} \frac{\del^2}{\del \tilde{\tau}^2} \tilde \theta_1 + \tilde v \pqty*{1 - \tilde v^2 }  \tilde \theta_1'' + \pqty*{2 - 5\tilde v^2} \tilde \theta_1' = \frac{12 h_1 i}{5 c_0}  \tilde v^4 \sinh(\omega \tilde \tau)  ,
\end{equation}
where \(\tilde \theta_1' = \del_{\tilde{v}} \tilde \theta_1\).  This
is a linear inhomogeneous \ac{pde}.  The standard approach is to look
for a particular solution \(\tilde \theta_{1,p}\) and then solve the
associated homogeneous equation to find the general solution to the
problem.

One can use separation of variables, observing that the inhomogeneity is proportional to \(\sinh(\omega \tilde \tau)\).
It follows that
\begin{equation}
  \tilde \theta_{1,p}(\tilde \tau, \tilde v ) = i\frac{h_1}{c_0}\sinh(\omega \tilde \tau) B_1(\tilde v)
\end{equation}
which leads to an inhomogeneous \ac{ode} for \(B_1^p(\tilde v)\):
\begin{equation}
      \tilde v \pqty*{1 - \tilde v^2 }  B_1'' + \pqty*{2 - 5\tilde v^2} B_1' + 3 \tilde v B_1 = \frac{12 i}{5 } \tilde v^4   .
\end{equation}
The source is an elementary function of \(\tilde v\), and therefore a standard variation of parameters\footnote{This method is reviewed in Appendix~\ref{sec:variation-parameters}.} leads to
\begin{equation}
  B_1(\tilde v) = - \frac{2}{15} \pqty*{\tilde v^3 + 6 \tilde v - \frac{2}{\tilde v}} .  
\end{equation}

Now that the particular solution is in hand, one is left with solving the associated homogeneous problem:
\begin{equation}
   \frac{3 \tilde v}{\omega^2} \frac{\del^2}{\del\tilde{\tau}^2} \tilde \theta_{h} + \tilde v \pqty*{1 - \tilde v^2 }  \tilde \theta_{h}'' + \pqty*{2 - 5\tilde v^2 }  \tilde \theta_{h}'  = 0 .
\end{equation}
Again searching for a solution by separating the variables,
\begin{equation}
  \tilde \theta_{h}(\tilde \tau, \tilde v) = A(\omega\tilde \tau) C(\tilde v),
\end{equation}
the homogeneous equation becomes
\begin{equation}
  \frac{\ddot A(\omega \tilde \tau)}{A(\omega \tilde \tau)} = \frac{\pqty*{5\tilde v^2 -2} C'(\tilde v) + \tilde v \pqty*{\tilde v^2 - 1 }  C''(\tilde v) }{3 \tilde v C(\tilde v)}  
\end{equation}
which admits the solution
\begin{align}
  &A(\omega \tilde \tau) = c_1 e^{\lambda \omega \tilde \tau} + c_2 e^{-\lambda \omega \tilde \tau} ,\\
  &\begin{aligned}
    C(\tilde v) &= k_1 \frac{\cos*(\sqrt{4 +3 \lambda^2} \arccos(\tilde v) )}{\tilde v \sqrt{1 - \tilde v^2}} + k_2 \frac{\sin*( \sqrt{4 + 3 \lambda^2} \arccos(\tilde v))}{\tilde v \sqrt{ 1 - \tilde v^2}} \\ &= \frac{k_1 \cos*(\sqrt{4 + 3 \lambda^2} \tilde \psi) + k_2 \sin*(\sqrt{4 + 3 \lambda^2} \tilde \psi) }{\sin(\tilde \psi) \cos(\tilde \psi)},
  \end{aligned}
\end{align}
where the last step has implemented the change of variables described above.

Putting the two elements together leads to the general solution of the problem:
\begin{multline}
  \tilde \theta_1(\tilde \tau, \tilde v) = - \frac{2 h_1 i}{15 c_0} \pqty*{\tilde v^3 + 6 \tilde v - \frac{2}{\tilde v}} \sinh(\omega \tilde \tau) + \pqty*{c_1 e^{\lambda \omega \tilde \tau} + c_2 e^{-\lambda \omega \tilde \tau}} \\
  \pqty*{k_1 \frac{\cos*(\sqrt{4 +3 \lambda^2} \tilde \psi )}{\sin(\tilde \psi)\cos(\tilde \psi)} + k_2 \frac{\sin*(\sqrt{4 + 3 \lambda^2} \tilde \psi)}{\sin(\tilde \psi) \cos(\tilde \psi)}} .
\end{multline}

The next step is the imposition of boundary conditions.  First, note that the
system has opposite insertions at \(\tilde \tau = \pm \infty \),
which selects solutions that are odd under \(\tilde \tau \to - \tilde
\tau\), fixing \(c_1 = - c_2 = 1/2\) (for convenience). As for the
dependence on \(\tilde v\), the coefficient of the highest derivative
term in the \ac{pde} vanishes for \(\tilde v = 0\) and \(\tilde v =
1\). If follows that the general solution is singular both in \(\tilde
v = 0\) and \(\tilde v = 1\). For \(\tilde \psi \sim 0 \) (center of
the droplet \(\tilde v=1\)), one has
\begin{equation}
   \tilde \theta_1(\tilde \tau, \tilde v ) \isEquivTo_{\tilde \psi \to 0} \frac{k_1}{\tilde \psi} \sinh(\lambda \omega \tilde \tau) + \text{regular} .
\end{equation}
There is no physical reason to expect a singularity at the center of
the droplet, and therefore \(k_1 = 0\).  At the edge, however, the
situation is different.  Already at \ac{lo}, before any perturbation
is considered, the large-charge approximation is not strictly valid
since the charge density goes to zero~\cite{Son:2005rv}. So there is
no reason to trust the solution at the droplet edge and impose
regularity.  Instead, the remaining parameters are fixed by minimizing
the energy and then considering in detail the consequences of the
(singular) behavior of \(\tilde X\) around the \ac{lo} edge \(\tilde
\psi = \psi_{\ac{lo}} = \pi/2\), in order to show that the
corresponding corrections are parametrically suppressed.

At leading order in \(r_0\), 
\begin{equation}
  \tilde X(\tilde \tau, \tilde \psi) = \mu \cos^2(\tilde \psi) - \frac{ r_0 \mu^{3/2} h_1}{c_0} B_1(\tilde v) \cosh(\omega \tilde \tau) + r_0 \mu^{3/2} k_2 \lambda \frac{\sin*(\sqrt{4 + 3 \lambda^2} \tilde \psi)}{\sin(\tilde \psi) \cos(\tilde \psi)} \cosh(\lambda\omega\tau) , 
\end{equation}
so that the Lagrange density at the saddle is
\begin{multline}
  \tilde {\cal L}_{\saddle} = -c_0 \mu^{5/2} \cos^5(\tilde \psi) - h_1 r_0 \mu^3 \cos^6(\tilde \psi) \cosh(\omega \tilde \tau) \\
  - \frac{5}{2} c_0 r_0 \mu^3 \cos^3(\tilde \psi) \pqty*{\frac{h_1 }{ c_0} B_1(\tilde v) \cosh(\omega \tilde \tau) +  k_2 \lambda \frac{\sin*(\sqrt{4 + 3 \lambda^2} \tilde \psi)}{\sin(\tilde \psi) \cos(\tilde \psi)} \cosh(\lambda \omega \tilde \tau)} .
\end{multline}
The action now takes the form
\begin{equation}
  \begin{aligned}
    \tilde S_{\saddle} &= \int_{-\infty}^{\infty} \dd{\tilde{\tau}} \int \dd^3{\tilde{x}} \tilde {\cal L}_{\saddle} = 4 \pi R_{\ac{lo}}^3 \int_{-T}^T \dd{\tilde{\tau}} \int_0^{\pi/2} \sin^2(\tilde \psi) \cos(\tilde \psi) \dd{\tilde{\psi}} \tilde {\cal L}_{\saddle} \\
    &= -\frac{5 \pi^2 c_0 T}{8 \sqrt{2}} \frac{\mu^4}{\omega^3} \Bigg[ 1 - h_1 r_0 \sqrt{\mu} \pqty*{\frac{31744}{4725}  \frac{\sinh(\omega T)}{\omega T c_0} + k_2 \frac{256 \sin(\frac{\pi}{2}\sqrt{4 + 3 \lambda^2})}{3 \pi \lambda^2 \pqty{4 -\lambda^2}} \frac{\sinh(\lambda \omega T)}{\omega T}} \Bigg],
  \end{aligned}
\end{equation}
which is minimized for \(k_2 = 0\). (Note that for arbitrary $\lambda >2$, in the continuation back to Minkowski space, the \(k_2 \) term correspond to higher harmonics which incur an energy cost.)

The final result is that the lowest-energy configuration at \ac{nlo} in the effective range in the oscillator frame is
\begin{equation}
  \tilde  \theta(\tilde \tau, \tilde v ) = - i \mu \tilde \tau + i \frac{h_1}{c_0} \frac{r_0  \mu^{3/2}}{\omega} B_1(\tilde v) \sinh(\omega \tilde \tau) .
\end{equation}

\subsection{Action at the saddle in the flat frame}
\label{sec:action-at-saddle-flat}

The next step is to move to the flat frame. The transformations of $\theta$ and the kinematical variables are given in Eq.~(\ref{eq:varchange}),
and
\begin{align}
  \tilde v^2 = 1 - \frac{\tilde r^2}{R_{\ac{lo}}^2} &\mapsto v^2 = 1 - \frac{r^2}{R_{\ac{lo}}^2 \pqty{1- \omega^2 \tau^2}} \\
  \sin(\tilde \psi) = \frac{\tilde r}{R_{\ac{lo}}} &\mapsto \sin(\psi) = \frac{r}{R_{\ac{lo}}\pqty{1 - \omega^2 \tau^2}^{1/2}}.
\end{align}
The volume form is then
\begin{multline}
  \label{eq:flat-frame-measure}
  \dd{\tau} \dd^3{x} = r^2 \dd{r} \dd{\tau} \dd{\Omega} = - R_{\ac{lo}}^3 \pqty{1 - \omega^2 \tau^2}^{3/2} \sqrt{ 1 - v^2} v \dd{v} \dd{\tau} \dd{\Omega} \\
  =  R_{\ac{lo}}^3 \pqty{1 - \omega^2 \tau^2}^{3/2} \sin^2(\psi) \cos(\psi) \dd{\psi} \dd{\tau} \dd{\Omega}.
\end{multline}
Finally, the solution in the flat frame is
\begin{equation}
  \theta(\tau, v) = \frac{i \mu}{2 \omega} \log \frac{1-\omega \tau}{1 + \omega \tau} + i \pqty*{1 - v^2} \mu \tau + i \frac{h_1}{c_0} r_0  \mu^{3/2} \frac{\tau}{\sqrt{ 1 - \omega^2 \tau^2 }} B_1( v)
\end{equation}
with the corresponding value of \(X\),
\begin{equation}
  X(\tau, v) = \frac{v^2 \mu}{ 1 - \omega^2 \tau^2} - r_0 \frac{h_1}{c_0} \pqty*{\frac{\mu}{1 - \omega^2 \tau^2}}^{3/2} B_1(v) .
\end{equation}
$X$ diverges at the droplet edge due to the divergence in $B_1$ which scales like $1/v$. This is, however, not an issue, as this divergence cancels out in the density. Expanding $\rho$ to leading order in $r_0$, one sees that
\begin{equation}
  \begin{aligned}
    \rho &= \frac{5}{2}c_0 X^{3/2} + 3 h_1 r_0 X^2\\
      &= \frac{5}{2}c_0 \pqty*{\frac{\mu}{1-\tau^2\omega^2}}^{3/2} v^3 +r_0 h_1 \pqty*{\frac{\mu}{1-\tau^2\omega^2}}^2 \pqty*{-\frac{15}{4}v B_1(v) + 3 v^4}.
  \end{aligned}
\end{equation}
Hence, at the droplet edge, \emph{i.e.} at $v=0$,
\begin{equation}
	\rho(v=0) = - r_0 h_1\pqty*{\frac{\mu}{1-\tau^2\omega^2}}^2,
\end{equation}
which is regular.

The computation of the action at the saddle parallels what was done in the previous section, giving the Lagrange density
\begin{equation}
  {\cal L}_{\saddle} = - c_0 \frac{\mu^{5/2} \cos^5(\psi)}{\pqty*{1 - \omega^2\tau^2}^{5/2}} - r_0 h_1 \pqty*{\frac{\mu}{1 - \omega^2 \tau^2}}^3 \pqty*{\cos^6(\psi) + \frac{5}{2} \cos^3(\psi) B_1(v)}
\end{equation}
and the action
\begin{equation}
  S_{\saddle} = - \frac{5\pi^2 c_0 \mu^4}{16 \sqrt{2} \omega^3} \int_{-1/\omega}^{1/\omega} \frac{\dd{\tau}}{1 - \omega^2 \tau^2} + \frac{1984 \pi \mu^{9/2} r_0 h_1}{945 \sqrt{2} \omega^3}  \int_{-1/\omega}^{1/\omega} \frac{\dd{\tau}}{\pqty*{1 - \omega^2 \tau^2}^{3/2}} \, .
\end{equation}
The integrals over \(\tau\) diverge. To regulate them, the integration bounds are restricted to \(\pm 1/\omega \mp \epsilon\) where finally one takes the \(\epsilon \to 0 \) limit.
The result is
\begin{equation}
  S_{\saddle} =  - \frac{5\pi^2 c_0 \mu^4}{16 \sqrt{2} \omega^4} \log*(\frac{2}{\epsilon \omega}) + \frac{1984 \pi^2 \mu^{9/2} r_0 h_1}{945  \omega^4} \frac{1}{(\omega \epsilon)^{1/2}} + \order{\epsilon^{1/2}} \, .
\end{equation}
The leading term agrees with the results of~\cite{Beane:2024kld}, given in Eq.~(\ref{eq:saddle-action}).

Now that the action at the saddle has been computed, it should be
inserted into the expression for the two-point function given in
Eq.~\eqref{eq:two-point-function}.  However, the \ac{nlo} correction
in \(r_0\) to the action at the saddle can be reabsorbed by a field
redefinition and is therefore not physical\footnote{The diverging
  integral can be regularized by analytic continuation, as is done in
  Section~\ref{sec:struct-solution-bulk}. Then, the NLO term vanishes.
  The coefficient in front of the log, on the other hand, is physical
  and appears as the residue of the pole in a ratio of gamma
  functions.}.  The final result is that there is no \ac{nlo}
correction in \(r_0\) to the two-point function. It is interesting
that the leading effective-range corrections are also found to vanish
in conformal perturbation theory in the three-body
sector~\cite{Chowdhury:2023ahp}.  In Ref.~\cite{Platter:2008cx} it was
shown that when computing the three-body binding energy in an \ac{eft} of contact
operators, there is a
discrete scale invariance which guarantees that the first-order effective-range
correction to the binding energy vanishes. Such a symmetry argument does
not seem to be available in the perturbation theory considered here and
in Ref.~\cite{Chowdhury:2023ahp}.

\subsection{Boundary effects}
\label{sec:boundary-effects}

In the \ac{nlo} calculation of the previous section, the fact that the
size of the droplet changes as the Schr\"odinger symmetry is deformed
was neglected.  Here it will be shown that the shift in the boundary
leads to corrections of order \(\order{r_0^{7/3}}\), which is higher
order than what is considered in this work.  It is important to remark
that in any case, close to the \ac{lo} edge, the solution that has
been found diverges and does not respect the conditions one expects to
be satisfied in a perturbative expansion.  For this reason one can at
best estimate the parametric dependence of the neglected terms.  A
similar situation arises also in the case of pure Schrödinger dynamics
at higher order, where the \ac{eft} requires the insertion of edge
operators~\cite{Son:2005rv,Hellerman:2020eff}.  In both cases, this near-edge
dynamics leads to new terms in the effective action at the saddle
point, scaling with rational powers of the expansion parameter (here
\(r_0 \sqrt{\mu}\)).

The boundary of the droplet is found by imposing \(X(\tau, \psi_{\ac{nlo}}) = 0\).
Explicitly:
\begin{equation}
  X(\tau, \psi_{\ac{nlo}}) = \frac{\mu}{1-\tau^2\omega^2} \cos^2(\psi_{\ac{nlo}}) + \pqty*{\frac{\mu}{1-\omega^2\tau^2}}^{3/2}r_0\frac{h_1}{c_0} B_1(\psi_{\ac{nlo}}) = 0 .
\end{equation}
As noted above, for \(\psi \to \pi/2\) the function \(B_1\) diverges as
\begin{equation}
  B_1(\psi ) \isEquivTo_{{\psi \to \pi/2}} \frac{4}{15\cos(\psi)}.
\end{equation}
It follows that \(\pi/2 - \psi_{\ac{nlo}}\) scales like \(r_0^{1/3}\):
\begin{equation}
  \cos(\psi_{\ac{nlo}}) = \pqty*{ \frac{4 r_0 h_1\sqrt{\mu}}{15 c_0 \sqrt{1 - \tau^2 \omega^2}}}^{1/3} \Rightarrow \psi_{\ac{nlo}} = \frac{\pi}{2} - \pqty*{ \frac{4 r_0 h_1\sqrt{\mu}}{15 c_0\sqrt{1 - \tau^2 \omega^2}}}^{1/3} .
\end{equation}
From this expression one can estimate the error that is made in computing the action at the saddle when integrating over \(\psi\) up to \(\psi = \pi/2\), in contrast to \(\psi_{\ac{nlo}}\):
\begin{equation}
  \Delta S \propto \frac{\mu^4}{\omega^3} \int \dd{\tau} \frac{1}{1 - \omega^2\tau^2 } \int_{\psi_{\ac{nlo}}}^{\pi/2} \sin^2(\psi) \cos(\psi) \dd{\psi} \cos^5(\psi) \propto \frac{\mu^4}{\omega^3} \int \dd{\tau}   \frac{ \pqty*{r_0 \sqrt{\mu}}^{7/3}}{\pqty*{1 - \omega^2\tau^2 }^{1+7/6}} .
\end{equation}
Unsurprisingly, the integral over \(\tau\) is divergent.  However, as
will be seen in the following, after regularization, there remains a
finite contribution to the action at the saddle.  For the moment, 
observe that such a contribution scales like \(r_0^{7/3}\) and is
parametrically smaller than the \ac{nnlo} term that one obtains by
adding a new term of order \(\order{r_0^2}\) to the action, and that
will be computed in the next section.

\subsection{Continuity equation and charge conservation}
\label{sec:continuity-charge}

The solution to the \ac{eom} has been computed in terms of the parameters \(r_0\), \(\omega = 2/\tau_{12}\) and \(\mu\).
However, while the first two parameters have well-defined physical meaning, in order to compute correlation functions
at fixed charge, it is necessary to express \(\mu\) as function of the inserted charge \(Q\).
To do so, consider the continuity equation
\begin{equation}
  \label{eq:continuity}
  i \pdv{\rho}{\tau} - \pdv*{\pqty*{\pdv{\theta}{x^i} \rho}}{x^i} = i Q \pqty*{ \delta(\tau + 1/\omega) - \delta(\tau - 1/\omega)} \delta(\mathbf{x}).
\end{equation}
Integrating this equation over the droplet will fix the charge. However, extra care is necessary for two reasons:
\begin{enumerate}
\item The position of the droplet edge varies over time (the Reynolds transport theorem);
\item The current \(j^i = \partial^i \theta \rho\) is not zero at the droplet edge because, while the density \(\rho\) vanishes, the term \(\partial^i\theta\) diverges.
\end{enumerate}
Taking these issues into account, the integral form is
\begin{equation}
  i \odv*{ \int_D \dd^3{x} \rho}{\tau} - i \int_{\del D} \rho \, \mathbf{v} \cdot \dd{\mathbf{\Sigma}} - \int_D \dd^3{x} \del_i(\del^i\theta \rho) = i Q \pqty*{ \delta(\tau + 1/\omega) - \delta(\tau - 1/\omega)} ,
\end{equation}
where \(\mathbf{v}\) is the Eulerian velocity of the edge.

In the radially-symmetric problem, the equation simplifies to
\begin{multline}
  i \odv*{\int_0^{R(\tau)} r^2 \dd{r} \rho(r)}{\tau} - i \eval*{r^2 \rho(r) \dot R(\tau)}_{r = R(\tau)} - \int_0^{R(\tau)} r^2 \dd{r} \pqty*{ \theta''(r) \rho(r) + \theta'(r) \rho(r) + \frac{2}{r} \theta'(r)\rho'(r)} \\
  = i \frac{Q}{4 \pi} \pqty*{ \delta(\tau + 1/\omega) - \delta(\tau - 1/\omega)} ,
\end{multline}
where \(R(\tau) = R_{\ac{lo}} \pqty*{1 - \omega^2 \tau^2}^{1/2} \) is the \ac{lo} edge.%
\footnote{For this argument it is sufficient to consider the edge at leading order. The NLO correction to the edge would reflect into higher-order corrections.}
Evaluating each term on the \ac{nlo} solution gives
\begin{allowdisplaybreaks}
  \begin{gather}
    \odv*{\int_0^{R(\tau)} r^2 \dd{r} \rho(r)}{\tau} = \frac{5\pi}{16 \sqrt{2}} \pqty*{\frac{\mu}{\omega}}^3 \pqty*{ \delta(\tau + 1/\omega) - \delta(\tau - 1/\omega)} - \frac{\sqrt{2}}{6} \frac{h_1 r_0 \sqrt{\omega}}{\sqrt{1- \omega^2 \tau^2}} \pqty*{\frac{\mu}{\omega}}^{7/2}  \\
    \eval*{r^2 \rho(r) \dot R(\tau)}_{r = R(\tau)} = - \frac{\sqrt{2}}{2} \frac{h_1 r_0 \sqrt{\omega}}{\sqrt{1- \omega^2 \tau^2}} \pqty*{\frac{\mu}{\omega}}^{7/2} ,\\
    \int_0^{R(\tau)} r^2 \dd{r} \pqty*{ \theta''(r) \rho(r) + \theta'(r) \rho(r) + \frac{2}{r} \theta'(r)\rho'(r)} = \frac{\sqrt{2}}{3}i \frac{h_1 r_0 \sqrt{\omega}}{\sqrt{1- \omega^2 \tau^2}} \pqty*{\frac{\mu}{\omega}}^{7/2} .
  \end{gather}
\end{allowdisplaybreaks}
The \(\order{r_0}\) terms cancel among each other and no time-independent \(\order{r_0}\) term remains.
The continuity equation then reduces to
\begin{equation}
  \frac{5 c_0 \pi^2}{4 \sqrt{2}} \pqty*{\frac{\mu}{\omega}}^3 = Q 
\end{equation}
or, equivalently, using the Bertsch parameter,
\begin{equation}\label{eq:mu-and-Q}
  \pqty*{\frac{\mu}{\omega}}^3 = 3 \xi^{3/2} Q,
\end{equation}
in agreement with Eq.~(\ref{eq:swseft27}).
This fixes the value of \(\mu\) as a function of \(Q\) in agreement with the result in~\cite{Beane:2024kld}, where this equation had been interpreted as charge conservation.

While this calculation has only computed the \(\order{r_0}\) term, the same cancellation must occur at all orders.
As discussed in the previous section, each power of \(r_0\) in the expression of the density \(\rho\) is accompanied by powers of \(\pqty{1 - \omega^2 \tau^2}\).
All of these terms have to cancel separately on the \ac{lhs} of the differential form of the continuity equation because there is no such \(\tau\) dependence on the \ac{rhs}.
The conclusion is that the expression of \(\mu\) as function of \(Q\) is valid to all orders in \(r_0\).

\section{Second-order effective range correction}
\label{sec:nnlo-range}

\subsection{EOM in the oscillator frame}
\label{sec:eom-oscillator-frame-nnlo}

Consider now the solution of the \ac{eom} for the action
\begin{equation}
  L = -c_0 X^{5/2} - h_1 r X^3 -  h_2\,r^2 X^{7/2}. 
\end{equation}
Following the same  procedure  as above, the solution at
order $\order{r_0^2}$ in the oscillator frame can be written as
\begin{equation}
	\tilde\theta(\tilde\tau,\tilde v) = \tilde \theta_0(\tilde\tau,\tilde v) +  \frac{\mu^{3/2}}{\omega} r_0 \tilde\theta_1(\tilde\tau,\tilde v) +  \frac{\mu^2}{\omega} r_0^2 \tilde\theta_2(\tilde\tau,\tilde v).
\end{equation}

The key observation is that even though the \ac{eom} is not separable, it becomes so order by order and one can decompose
\begin{equation}
	\tilde\theta_2 (\tilde \tau, \tilde v) = i\sinh(2 \omega \tilde \tau) B_2(\tilde v),
\end{equation}
where, in analogy with the first-order case,  only the term that survives after minimizing the energy has been given.
The second-order \ac{eom} becomes
\begin{gather}
	\tilde v(1-\tilde v^2) B_2''(\tilde v) + (2-5\tilde v^2) B_2'(\tilde v) + 12 \tilde v B_2(\tilde v) = f_2(\tilde v),\\
	f_2(\tilde v) = \frac{210 c_0 h_2 \tilde v^{10} + 2 h_1^2 \pqty*{8 + 12 \tilde v^2 + 22 \tilde v^4 - 36 \tilde v^6 + 9 \tilde v^8 - 64 \tilde v^{10}}}{75 c_0^2 \tilde v^5}.
\end{gather}
This equation can be solved explicitly using variation of parameters since the associated homogeneous equation is hypergeometric. 
It admits the solution
\begin{equation}
	C_2(\tilde v) = k_1 \frac{1-8\tilde v^2+8\tilde v^4}{\tilde v\sqrt{1-\tilde v^2}} + k_2 \pqty*{1-2\tilde v^2}.
\end{equation}
The integration constant $k_1$ is fixed to $0$ by requiring the
solution to be regular in the center of the droplet $\tilde v=1$, so the
associated solution just yields one term involving a constant $k_2$ that is
fixed to zero by minimization of the energy.  The final result is
\begin{allowdisplaybreaks}
  \begin{multline}
    B_2(\tilde \psi) = B_2^2 \frac{h_2}{c_0} +B_2^1 \pqty*{\frac{h_1}{c_0}}^2 =\\
    - \frac{7}{4800 \sin(2 \tilde \psi)}\bqty*{ 60 \psi \cos(4 \tilde \psi) + 50 \sin(2 \tilde \psi) + 31 \sin(4 \tilde \psi) - 6 \sin(6 \tilde \psi) } \frac{h_2}{c_0}\\
    + \Bigg[ \frac{384 \cos(4 \tilde{\psi}) - 26609 \cos(2 \tilde{\psi}) - 8728}{36000} + \frac{4}{15} \cos(2 \tilde{\psi}) \log(\cos(\tilde{\psi})) \\
    - \frac{7 \psi \cos(4 \tilde{\psi})}{80 \sin(2 \tilde{\psi})} + \frac{4}{75 \cos^2(\tilde{\psi})} +  \frac{4}{225 \cos^4(\tilde{\psi})} \Bigg] \pqty*{\frac{h_1}{c_0}}^2.
  \end{multline}
\end{allowdisplaybreaks}
The inhomogeneous solution diverges at the boundary of the droplet $v=0$ as
\begin{equation}
	B_2(v) \xrightarrow[v\to 0]{} \pqty*{\frac{2 h_1}{15 c_0}}^2 \frac{1}{v^4} + \frac{1}{3} \pqty*{\frac{2 h_1}{5 c_0}}^2 \frac{1}{v^2} - \frac{7\pi}{320 c_0^2}\frac{h_1^2+c_0 h_2}{v} - \frac{4}{15}\pqty*{\frac{h_1}{c_0}}^2 \log v.
\end{equation}
Recall that the \ac{nlo} solution was understood to be valid only up to a distance $\order{r_0^{2/3}}$ from $v=0$, so the singularity in $v=0$ is not a problem.

\subsection{Action at the saddle in the flat frame}
\label{sec:action-at-saddle-flat-nnlo}

Reverting to the flat frame gives:
\begin{multline}
  X = \frac{\mu}{1 - \omega^2 \tau^2} \cos^2(\psi) - \frac{h_1}{c_0} \pqty*{\frac{\mu}{1 - \omega^2 \tau^2}}^{3/2} r_0 B_1(\psi) \\
    - \pqty*{\frac{\mu}{1 - \omega^2 \tau^2}}^2 \frac{r_0^2}{4} \Bigg[ 8 \frac{h_2}{c_0} \pqty{ 1 + \omega^2 \tau^2} B_{2}^{(2)}(\psi) \\ + \pqty*{\frac{h_1}{c_0}}^2 \pqty*{ 8 \pqty{1 + \omega^2 \tau^2} B_2^{(1)}(\psi) - \tau^2 \omega^2 \tan^2(\psi) B_1'(\psi)^2 }  \Bigg].
\end{multline} 
As expected, each order in $r_0$ comes with an extra $\pqty{1-t^2\omega^2}^{1/2}$ in the denominator.
The Lagrange density at the saddle is
\begin{equation}
  {\cal L}_{\saddle} = - c_0 \pqty*{\frac{\mu}{1 - \omega^2 \tau^2}}^{5/2} L_0 -  r_0 \pqty*{\frac{\mu}{1 - \omega^2 \tau^2}}^3 h_1 L_1 - r_0^2 \pqty*{\frac{\mu}{1 - \omega^2 \tau^2}}^{7/2} \pqty*{\frac{h_1^2}{c_0} L_2^{(1)} + h_2 L_2^{(2)}}
\end{equation}
with
\begin{align}
  L_0 &= \cos^5(\psi),  \\
  L_1 &= \frac{5}{2} B_1(\psi) \cos^3(\psi) - \cos^6(\psi), \\
  L_2^{(1)} &=
              \begin{multlined}[t][]
                \frac{15}{8} B_1(\psi)^2 \cos(\psi) - 5 \pqty{ 1 + \omega^2 \tau^2} B_2^{(1)}(\psi) \cos^3(\psi) - 3 B_1(\psi) \cos^4(\psi) \\ + \frac{5}{8} \tau^2 \omega^2 B_1'(\psi)^2 \cos(\psi) \sin^2(\psi) ,
              \end{multlined}
  \\
  L_2^{(2)} &= \cos^7(\psi) - 5 \pqty{1 + \omega^2 \tau^2} B_2^{(2)}(\psi) \cos^3(\psi).
\end{align}
Now one proceeds as before to the action at the saddle point.%
\footnote{The integration bounds for \(\psi\) are again \(0\) and \(\pi/2\), up to higher-order corrections.}
The \(\order{r_0^2}\) term comprises four corrections:
\begin{equation}
  \eval*{S_{\saddle}}_{r_0^2} = r_0^2 \frac{\mu^5}{\omega^3} \int_{-1/\omega}^{1/\omega} \dd{\tau} \frac{h_1^2/c_0 \pqty*{ s_{2,1} + s_{2,2} \omega^2 \tau^2} + h_2 \pqty*{ s_{2,3} + s_{2,4} \omega^2 \tau^2}}{\pqty*{1- \omega^2 \tau^2}^2}  
\end{equation}
with
\begin{align}
  \label{eq:s2i}
  s_{2,1} &= \frac{13787 \pi^2}{12800 \sqrt{2}} \!+\! \frac{\pi^2 \log(2)}{6 \sqrt{2}}, &
  s_{2,2} &= \frac{48281 \pi^2}{38400 \sqrt{2}} \!+\! \frac{\pi^2 \log(2)}{6 \sqrt{2}}, &
  s_{2,3} &= -\frac{273 \pi^2}{2500 \sqrt{2}}, &
  s_{2,4} &= -\frac{833 \pi^2}{2560 \sqrt{2}}.
\end{align}
Regularizing the integral over \(\tau\), one again finds a term that diverges
as \(1/\epsilon\) and can be reabsorbed in the definition of the
operators, and a logarithmic divergence, whose coefficient is the
physical quantity of interest:
\begin{multline}
  \eval*{S_{\saddle}}_{r_0^2} = \frac{r_0^2 \mu^5}{2 \omega^4} \pqty*{\pqty{s_{2,1} - s_{2,2}} \frac{h_1^2}{c_0} + \pqty{s_{2,3} - s_{2,4}} h_2} \log*(\frac{2}{\omega\epsilon}) \\
  = \frac{r_0^2 \mu^5}{ \omega^4} \pqty*{ - \frac{173 \pi^2}{1920 \sqrt{2}} \frac{h_1^2}{c_0} + \frac{7 \pi^2}{64 \sqrt{2}} h_2} \log*(\frac{2}{\omega\epsilon}). 
\end{multline}
This is the final result for the correction of order \(\order{r_0^2}\) to the effective action and to the two-point function.
As expected, the numerical coefficients are of order one.
\begin{align}
     - \frac{173 \pi^2}{1920 \sqrt{2}} &\approx -0.629, & \frac{7 \pi^2}{64 \sqrt{2}} &\approx 0.763 .
\end{align}

\section{Regularization of the two-point function}
\label{sec:regularization2pt}

This section considers the regularization of the two-point function at the temporal boundary in some detail.
The two-point function at $\order{r_0^2}$ is
\begin{equation}
	G_Q(-1/\omega, 1/\omega) = \mathcal{O}_{\Delta,Q}(-1/\omega) \mathcal{O}_{\Delta,-Q}(1/\omega)e^{-S_{\saddle}}.
\end{equation}
Each of the three factors on the right-hand side of this equation are
divergent. The mutual cancellation of the divergences will give the
final result. The operator $\mathcal{O}_{\Delta,Q}$ is expressed in
terms of $\theta$ by identifying its charge and operator dimension and
is given by Eq.~(\ref{eq:swseft2}). As all divergences should be accounted for,
it is convenient to rewrite the correlation function in the form of
an integral over $\tau$, expressing the insertions at the boundary as
integrals of total derivatives:
\begin{align}
	\mathcal{O}_{\Delta,Q}(-1/\omega) \mathcal{O}_{\Delta,-Q}(1/\omega) &= \mathcal{N}^2 (X(-1/\omega)X(1/\omega))^{\Delta/2}e^{iQ(\theta(1/\omega)-\theta(-1/\omega))}\\
	&= \mathcal{N}^2  X(0)^\Delta \exp\pqty*{\int_{-1/\omega}^{1/\omega}\dd{\tau}\frac{\Delta}{2}\abs{\del_\tau\log X(\tau)}+iQ\del_\tau\theta(\tau)}.
\end{align}

Above it was found that  at \ac{nnlo},
\begin{multline}
	S_{\saddle} = Q^{4/3} \omega \int_{-1/\omega}^{1/\omega}\dd{\tau} \left(s_0 \frac{1}{1-\omega^2\tau^2}+h_1s_1\kappa \frac{1}{(1-\omega^2\tau^2)^{3/2}}\right.\\ 
	+ \left.\frac{\kappa^2}{(1-\omega^2\tau^2)^2}\pqty*{\frac{h_1}{c_0}(s_{2,1}+s_{2,2}\omega^2\tau^2 ) + h_2(s_{2,3}+s_{2,4}\omega^2\tau^2 )}\right),
\end{multline}
\begin{align}
	\del_\tau \theta &= i\gamma \frac{\omega}{1-\omega^2\tau^2} + i\frac{h_1}{c_0}\gamma\omega \frac{B_1(1)\kappa}{(1-\omega^2\tau^2)^{3/2}}+\frac{2i\gamma \omega (1-\omega^2\tau^2) B_2(1)\kappa^2}{(1-\omega^2\tau^2)^2}\\	\del_\tau \log X &= \frac{2\tau\omega}{1-\omega^2\tau^2} - \frac{h_1}{c_0}\gamma\omega^2 \frac{\tau B_1(1)\kappa}{(1-\omega^2\tau^2)^{3/2}} + \dots ,
\end{align}
with \(s_0 = 5 \pi^2 \sqrt{2} c_0/32\), \(s_1 = 1984 \pi \sqrt{2}/1890 \) and the \(s_{2,i}\) as defined in Eq.~\eqref{eq:s2i}.
Observe that at each order the three terms have the same analytic structure in terms of poles and branch cuts at the extrema of integration $\tau=\pm 1/\omega$.

All the integrals are divergent and require regularization. One possibility is to change the boundaries of integration,
\begin{equation}
	\int_{-1}^1 \dd{z} \frac{1}{(1-z^2)^n} = \lim_{\epsilon \to 0} \int_{-1+\epsilon}^{1-\epsilon} \dd{z} \frac{1}{(1-z^2)^n}.
\end{equation}
Another possibility is to analytically continue the power appearing in the denominator:
\begin{equation}
	\int_{-1}^1 \dd{z} \frac{1}{(1-z^2)^n} = \lim_{\delta \to 0} \int_{-1}^{1} \dd{z} \frac{1}{(1-z^2)^{n+\delta}}.
\end{equation}
The two regularizations are discussed in detail for these integrals and a generalization in Appendix~\ref{sec:struct-solution-bulk}.

\paragraph{\Acl{lo}.}
Using the two regularization schemes, the leading-order result is 
\begin{equation}
  \begin{aligned}
    \int_{-1}^1 \dd{z} \frac{s_0 Q^{4/3}-\gamma Q+\Delta\abs{z}}{(1-z^2)} &= \lim_{\epsilon \to 0}(-\Delta +Q\gamma -s_0 Q^{4/3})\log\epsilon -(-\Delta +Q\gamma - s_0 Q^{4/3})\log 2\\
                                                     &= \lim_{\delta\to 0} \frac{-\Delta +Q\gamma - s_0 Q^{4/3}}{\delta} +(\gamma Q - s_0 Q^{4/3})\log 2.
  \end{aligned}
\end{equation}
In both cases, the result is divergent unless
\begin{equation}
  \label{eq:Legendre-LO}
	-\Delta +Q\gamma - s_0 Q^{4/3} = 0.
\end{equation}
This is, as already observed in~\cite{Beane:2024kld}, the Legendre transform relating $\Delta$ to $s_0$.
The final result is
\begin{equation}
	G_Q(-1/\omega, 1/\omega) \propto  \frac{(2\gamma)^\Delta}{\tau_{12}^\Delta},
\end{equation}
as predicted by Schr\"odinger invariance.

\paragraph{\Acl{nlo}.}
At \ac{nlo}, in the $\delta$ regularization, the three contributions vanish separately since for the \(S\) and \(\theta\) part,
\begin{equation}
  \lim_{\delta\to 0}\int_{-1}^1 \dd{z} \frac{1}{(1-z^2)^{3/2}} =  \lim_{\delta\to 0} \sqrt{\pi} \frac{\Gamma(-1/2-\delta)}{\Gamma(-\delta)} = 0 .
\end{equation}
In the $\epsilon$ regularization, there remains a divergent contribution
\begin{equation}
	\lim_{\epsilon\to 0} \int_{-1+\epsilon}^{1-\epsilon} \dd{z} \frac{1}{(1-z^2)^{3/2}} = \lim_{\epsilon\to 0} \sqrt{\frac{2}{\epsilon}}.
\end{equation}
This term is however constant in the sense that it does not depend on $1/{\tau_{12}}=\omega$ and can be absorbed in the normalization of the operators in the two-point function.

As for the contribution from the integral of $\abs{\del_\tau \log X}$, this cancels with $\exp(\Delta X_{NLO}(0))$.
The regularization of the integral gives
\begin{equation}
  \lim_{\delta\to 0}\int_{-1}^1 \dd{z} \frac{\abs{z}}{(1-z^2)^{3/2+\delta}} =  \lim_{\delta\to 0}\frac{1}{-1/2-\delta} = -2,    
\end{equation}
and so the contribution to the two-point function is
\begin{multline}
  X(0)^{\Delta}_{\ac{nlo}} \exp*[\frac{\Delta}{2}\int_{-1/\tau}^{1/\tau} \dd{\tau} \abs*{\del_\tau \log(X)_{\ac{nlo}}}] =\\
  \exp*[-\Delta \frac{h_1}{c_0}\gamma\kappa B_1(1) ] \exp*[ \lim_{\delta\to 0}\pqty*{-\frac{\Delta h_1}{2 c_0}\gamma \kappa B_1(1)\int_{-1}^1 \dd{z} \frac{1}{(1-z^2)^{3/2-\delta}} }] = 1.
\end{multline}

The result is then that \emph{the \ac{nlo} correction in \(r_0\) is identically zero}.

\paragraph{\Acl{nnlo}.}

Using the same argument as above, one finds that the \(X^{\Delta}\) contribution vanishes identically at \ac{nnlo}.
It is convenient to collect the remaining \ac{nnlo} terms in the form:
\begin{multline}
  \pqty*{-2 \gamma Q B_2(1)  + \pqty*{\frac{h_1}{c_0} s_{2,2} + h_2 s_{2,4}}} \int_{-1}^1 \dd{z} \frac{1 + z^2}{\pqty{1 - z^2}^2} \\
  + \pqty*{ \frac{h_1}{c_0} \pqty{s_{2,1} - s_{2,2}} + h_2 \pqty{s_{2,3} - s_{2,4}}} \int_{-1}^1 \dd{z} \frac{1}{\pqty{1 - z^2}^2} .   
\end{multline}
Both integrals need to be regularized, and one finds, respectively,
\begin{equation}
  \begin{aligned}
    \int_{-1}^1 \dd{z} \frac{1 + z^2}{\pqty{1 - z^2}^{2}} &= \lim_{\delta \to 0}\frac{\sqrt{\pi} \Gamma(-\delta-1)}{\Gamma(-1/2-\delta)} = -1 \\
    &= \lim_{\epsilon\to 0} \pqty*{\frac{1}{\epsilon} } - \frac{1}{2}
  \end{aligned}
\end{equation}
and
\begin{equation}
  \begin{aligned}
    \int_{-1}^1 \dd{z} \frac{1 }{\pqty{1 - z^2}^{2}} &= \lim_{\delta \to 0} \pqty*{- \frac{1}{2 \delta}} - \frac{1}{2} + \log(2) \\
    &= \lim_{\epsilon \to 0} \pqty*{\frac{1}{2 \epsilon} - \frac{1}{2} \log(\epsilon)} + \frac{1}{2} \log(2) - \frac{1}{4}.
  \end{aligned}
\end{equation}
The  \(1/\delta\) and \(\log(\epsilon)\) divergences come again with the same coefficient.
Their cancellation results in the \(\order{\kappa^2}\) correction to the Legendre transform in Eq.~\eqref{eq:Legendre-LO}:

\begin{equation}%
\label{eq:Legendre--NNLO}
  \Delta = Q\gamma - s_0 Q^{4/3} - \frac{\kappa^2 Q^{4/3}}{2} \pqty*{ \frac{h_1^2}{c_0} \pqty{s_{2,1} - s_{2,2}} + h_2 \pqty{s_{2,3} - s_{2,4}}}.
\end{equation}
The coefficient in front of the integrals does not depend on \(\omega =
2/\tau_{12}\), and all the other (scheme-dependent) terms can be
treated as constants to be absorbed in the normalization of the
operators.  Note that the \ac{nnlo} corrections to the insertions do
not influence the form or the value of the two-point function.

The final result is once more
\begin{equation}
	G_Q(-1/\omega, 1/\omega) \propto  \frac{1}{\tau_{12}^\Delta},
\end{equation}
as expected.

\section{Leading-order solution for the scattering length corrections}
\label{sec:losls}

The same technique that was used to analyze the effective-range
corrections can be used to study the effect of the breaking of
Schr\"odinger invariance due to a finite scattering length\footnote{These effects were studied in Ref.~\cite{Beane:2024kld}. Here
it is shown that the results obtained in that paper are valid in the presence of boundary terms.}. In this
case, the  Lagrange density is
\begin{equation}
	{\cal L} = -c_0 X^{5/2} -{g_1}{a^{-1}} X^2 + \order{a^{-2}}.
\end{equation}
In the oscillator frame, 
\begin{equation}
	\tilde {\cal L} = -c_0 \tilde X^{5/2} - \frac{g_1}{a \cosh(\omega\tilde\tau)} \tilde X^2.
\end{equation}
One can search for isotropic solutions of the form
\begin{equation}
	\tilde\theta(\tilde\tau,\tilde v) = -i\mu \tilde\tau + i\frac{\mu^{1/2}}{\omega}\frac{g_1}{a}\tilde\theta_1(\tilde\tau,\tilde v).
\end{equation}
The \ac{nlo} \ac{eom} is then found to be
\begin{equation}
	3\tilde v \frac{\del^2}{\del \tilde{\tau}^2} \tilde\theta_1 +\omega^2 \tilde v(1-\tilde v^2)\tilde\theta_1''+(2-5\tilde v^2)\omega^2\tilde\theta_1' = -\frac{8}{5c_0} \tilde v^2 \omega \odv*{\frac{1}{\cosh(\omega\tilde\tau)}}{\tilde{\tau}} ,
\end{equation}
which is the same as Eq.~(\ref{eq:inho-r}), but with a different \ac{rhs}.
This equation is inhomogeneous and does not admit a solution by simple separation of variables. Consider, however, the ansatz
\begin{equation}
	\tilde\theta_1(\tilde\tau,\tilde v) = \frac{1}{\tilde v} \phi_1(\tilde\tau) + \tilde v \phi_2(\tilde v),
\end{equation}
which reduces the \ac{eom} to a system of \acp{ode} for $\phi_1$ and $\phi_2$:
\begin{equation}
  \begin{cases}
    \ddot\phi_1(\tau) + \omega^2\phi_1(\tau) + \frac{2}{3}\omega^2 \phi_2(\tau) = 0,\\
    \ddot\phi_2(\tau) - \frac{5}{3}\omega^2\phi_2(\tau) + \frac{8\omega^2}{15c_0}\frac{\sinh(\omega\tau)}{\cosh^2(\omega\tau)} = 0.
  \end{cases}
\end{equation}
The associated homogeneous system is hypergeometric and can be solved explicitly, and once again variation of parameters may be used to obtain
\begin{align}
   \phi_1(\tilde \tau) &= 
                         \begin{multlined}[t]
                           \frac{2}{5 c_0 \pqty{3 + \sqrt{15}}} e^{\omega \tilde{\tau}} \pqty*{\pqty{4 + \sqrt{15}} \textstyle\pFq{2}{1}{1, \frac{3-\sqrt{15}}{6}}{\frac{9-\sqrt{15}}{6} }{-e^{-2 \omega \tilde{\tau}}}  - \textstyle\pFq{2}{1}{1, \frac{3+\sqrt{15}}{6}}{\frac{9+\sqrt{15}}{6} }{-e^{-2 \omega \tilde{\tau}}}  } \\
                           + \frac{\sqrt{2}}{15 c_0} e^{i \pi/4} e^{\omega \tilde{\tau}} \textstyle\pFq{2}{1}{\frac{1-i}{2}, 1}{\frac{3-i}{2}}{-e^{-2 \omega \tilde{\tau}}} + \text{c.c.},
                         \end{multlined} \\
  {\phi}_2(\tilde{\tau}) &= - \frac{4}{15 c_0} e^{\omega \tilde{\tau}} \pqty*{ \pqty{3 + \sqrt{15}} \textstyle\pFq{2}{1}{1, \frac{3 - \sqrt{15}}{6}}{\frac{9 - \sqrt{15}}{6}}{-e^{2 \omega \tilde{\tau}}} + \pqty{3 - \sqrt{15}} \textstyle\pFq{2}{1}{1, \frac{3 + \sqrt{15}}{6}}{\frac{9 + \sqrt{15}}{6}}{-e^{2 \omega \tilde{\tau}}} }.
\end{align}
These explicit expressions are exact but complicated to use. A useful observation is that $\phi_1$ and $\phi_2$ admit an expansion in hyperbolic sines of odd multiples of $\omega\tilde\tau$:
\begin{align}
	\phi_1 &= \frac{1}{c_0}\sum_{n=0} \varphi_{1,n}\sinh*(\pqty{2n + 1} \omega \tau), &
	\phi_2 &= \frac{1}{c_0}\sum_{n=0} \varphi_{2,n}\sinh*(\pqty{2n + 1} \omega\tau).
\end{align}
The first few coefficients are given in Table~\ref{tab:coeff}.
\begin{table}
\centering
	\begin{tabular}{ccccc} \toprule
		$n$ & 0 & 1 & 2 & 3\\ \midrule
		$\varphi_{1,n}$ & $-\sfrac{4}{15}$ & $0$ & $0$ & $-\sfrac{2}{4725}$\\
		$\varphi_{2,n}$ & $\sfrac{4}{15}$ & $0$ & $\sfrac{2}{75}$ & $-\sfrac{26}{945}$\\ \bottomrule
	\end{tabular}	
	\caption{First coefficients in the expansions of $\phi_{1,2}$}
	\label{tab:coeff}
\end{table}
With this observation, one reverts to the flat frame in which each
$\sinh((2n+1)\omega\tilde\tau)$ turns into an odd power of
$(1-\omega^2\tau^2 )$ as in Eq.~\eqref{eq:sinh-arctanh}.  Following
the same construction as in Appendix~\ref{sec:struct-solution-bulk}, one
concludes that all of these terms do not contribute to the action at
the saddle, in analogy to all the $\order{r_0}$ corrections of odd
order.  The final result is that the only correction at $\order{1/a}$
is due to the evaluation of the \ac{nlo} action on the \ac{lo}
solution in agreement with the result in~\cite{Beane:2024kld}:
\begin{equation}\label{eq:sspa}
	\eval*{S_{\saddle}}_{a^{-1}} = \frac{64 \times 3^{1/6} \sqrt{2} \pi^2 \xi^{7/4}g_1}{35 a \omega^{1/2}} Q^{7/2}.
\end{equation}

\section{Minkowski-space correlation functions}
\label{sec:mscf}

\subsection{General procedure}

In order to make contact with phenomena, the position-space correlation functions
should be continued from Euclidean space back to Minkowski space, with the choice
of source points $x_1=(t,{\mathbf x})$, $x_2=(0,{\mathbf 0})$, and then, the
resulting correlation function, $G_Q(x_1;x_2)$ should be Fourier transformed to
obtain the momentum-space correlation function $G_Q(E,{\mathbf p})$, which is
then relevant to the propagation of matter in spacetime~\cite{Hammer:2021zxb,Braaten:2023acw},
as will be seen below.

\subsection{Relevant deformations: the scattering length}

\begin{figure}[!ht]
\centering
\includegraphics[width = 0.9\textwidth]{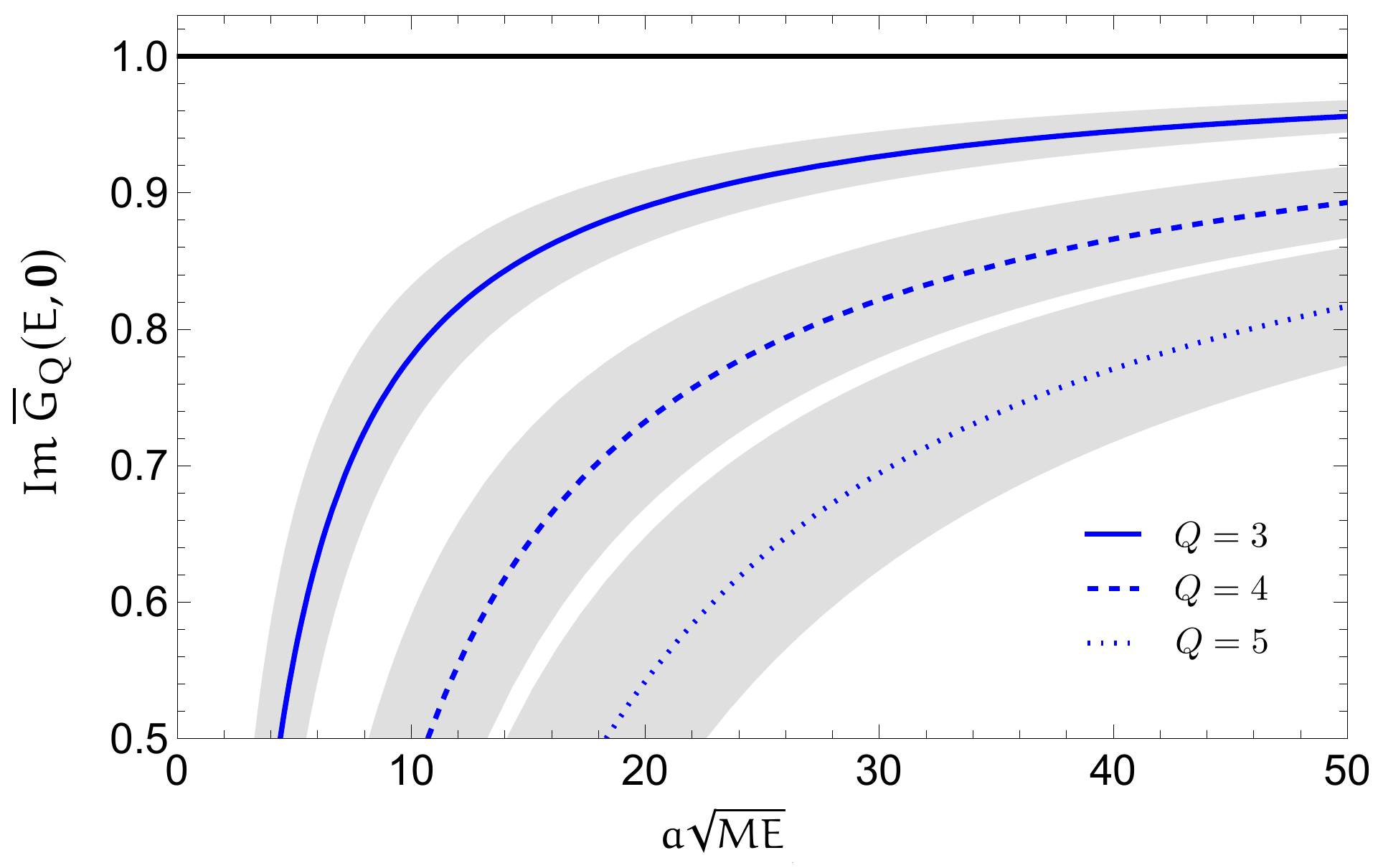}
\caption{Plot of the function $\Im{\overline G}_Q(E,{\mathbf 0})$ vs $a\sqrt{ME}$ ($a$ is assumed --without loss of generality-- to be negative). The solid, dashed, dotted blue
  curves correspond respectively to $Q=3$, $4$, $5$. The gray bands correspond to propagation of uncertainties in $g_1$.}    
  \label{fig:ImbarGa}
\end{figure}

On general grounds, one expects that the leading scattering length effects will take the form
\begin{eqnarray}
\Im G_Q(E,{\mathbf 0}) &=& C_0\;E^{\Delta_Q-5/2} \Bigg\lbrack 1  \ +\ \frac{\cal C_Q}{a\sqrt{ME}}\Bigg\rbrack\; ,
\label{eq:swscbe8}
\end{eqnarray}
where $C_0$ is a normalization constant that can be absorbed into the definition of the $X$ field.
In compact form, the symmetry-breaking action at the saddle point, Eq.~(\ref{eq:sspa}), in Euclidean space is
\begin{eqnarray}
\eval*{S_{\saddle}}_{a^{-1}}&=& \frac{64 \pi^2 g_1 \gamma^{7/2}}{105} a^{-1}\tau_{12}^{1/2}\ .
\label{eq:swscbe2}
\end{eqnarray}
As this contribution is independent of the regulator $\epsilon$, it can be
directly substituted into Eq.~(\ref{eq:cpt1}), which is then expanded
to leading order in $a^{-1}$. Following the procedure described above (using the Fourier transform given in Appendix~\ref{app:ft}) and
matching to Eq.~(\ref{eq:swscbe8}), one finds~\cite{Beane:2024kld}
\begin{eqnarray}
  {\cal C_Q}(Q)&=&  -\frac{64 \pi^2 g_1 \gamma^{7/2}}{105}
  \frac{\Gamma\left(\frac{6}{2}-\Delta_Q\right)}{\Gamma\left(\frac{5}{2}-\Delta_Q\right)}\tan\pi\Delta_Q .
  \label{eq:swscbe9}
\end{eqnarray}
It is noteworthy and promising that for $Q\sim 3$ these $\mathcal{O}(a^{-1})$ corrections in the large-charge \ac{eft} are consistent with the range of values found in
Ref.~\cite{Chowdhury:2023ahp} working directly with the three-body wavefunctions. 
To get a sense of the validity of perturbation theory it is convenient to define the function
\begin{eqnarray}
\Im{\overline G}_Q(E,{\mathbf 0}) &\equiv& \Im G_Q(E,{\mathbf 0})/\Im G^{CFT}_Q(E,{\mathbf 0}),
\label{eq:pttest}
\end{eqnarray}
where the \ac{cft} superscript indicates the correlation function evaluated with ${\cal C_Q}(Q)=0$.
This function is plotted in Fig.~\ref{fig:ImbarGa} vs $a\sqrt{ME}$ to illustrate the validity of the perturbative expansion.
Clearly perturbation theory works best at large scattering length and/or large energies. Note that the growth of 
${\cal C_Q}(Q)$ with $Q$ suggests that for fixed, large scattering length, as $Q$ is increased one must consider higher-energy propagation
to remain in the perturbative regime.

\subsection{Irrelevant deformations: effective range}

Here again on general grounds, one expects 
\begin{eqnarray}
\Im G_Q(E,{\mathbf 0}) &=& C_0\;E^{\Delta_Q-5/2} \Big\lbrack 1  \ +\ {\cal C'_Q}{ r_0 \sqrt{ME}}\Big\rbrack\; .
\label{eq:swscbe8b}
\end{eqnarray}
However, as shown above, the leading effective-range corrections vanish and therefore
\begin{eqnarray}
{\cal C'_Q}&=& 0 \ .
\end{eqnarray}
Note that this parallels what occurs at $Q=3$ in Ref.~\cite{Chowdhury:2023ahp}. (One easily
sees that perturbations at all odd powers of the range are power divergent with no finite part.)

Hence, the subleading effective-range corrections should enter as
\begin{eqnarray}
\Im G_Q(E,{\mathbf 0}) &=& C_0\;E^{\Delta_Q-5/2} \Big\lbrack 1  \ +\ {\cal C''_Q}{ r_0^2 {ME}}\Big\rbrack\; .
\label{eq:swscbe8c}
\end{eqnarray}
In compact form, the symmetry-breaking action at the saddle point, Eq.~(\ref{eq:sspr2}), in Euclidean space is
\begin{eqnarray}\label{eq:sspr2}
  \eval*{S_{\saddle}}_{r_0^2}  &=& r_0^2 \gamma^5 \pqty*{ - \frac{173\sqrt{2} \pi^2}{1920} \frac{h_1^2}{c_0} + \frac{7\sqrt{2} \pi^2}{64} h_2}\tau_{12}^{-1} \log*(\frac{\tau_{12}}{\epsilon}). 
\end{eqnarray}
It is convenient to define
\begin{eqnarray}
  \alpha\ \equiv\  \gamma^5 \pqty*{ - \frac{173\sqrt{2} \pi^2}{1920} \frac{h_1^2}{c_0} + \frac{7\sqrt{2} \pi^2}{64} h_2},
\end{eqnarray}
where the values of $h_{1,2}$ and $c_0$ extracted from lattice \ac{mc} data were given in Eq.~\eqref{eq:g0g1h1}.
Then, following the procedure outlined above, the conformal dimension in the presence of the symmetry breaking is
\begin{equation}
 \Delta = \Delta_Q \ +\ r_0^2\alpha \tau_{12}^{-1}\ .
\end{equation}
It is not particularly surprising that the conformal dimension is spacetime dependent in the
absence of Schr\"odinger symmetry. This then leads to the solution
\begin{equation}
G(x_1,x_2) \ = \  G_{CFT}(x_1,x_2) \tau_{12}^{-r_0^2\alpha \tau_{12}^{-1}} \ .
\label{eq:cptg}
\end{equation}   
To leading order in $\alpha$ and keeping terms of order $\order{\tau_{12}^{-1}\log\tau_{12}}$,
\begin{equation}
G(x_1,x_2) \ = \  G_{CFT}(x_1,x_2) \big\lbrack 1 - r_0^2\alpha \tau_{12}^{-1}\log\left(\tau_{12}\lambda \right) \big\rbrack \ ,
\label{eq:cptg2}
\end{equation}   
where $\lambda$ is an arbitrary energy scale. As the sole momentum
scale is $r_0^{-1}$, one expects $\lambda^{-1}\propto M r_0^2$. 
The constant of proportionality will be chosen below to optimize
perturbation theory.

\begin{figure}[!ht]
\centering
\includegraphics[width = 0.9\textwidth]{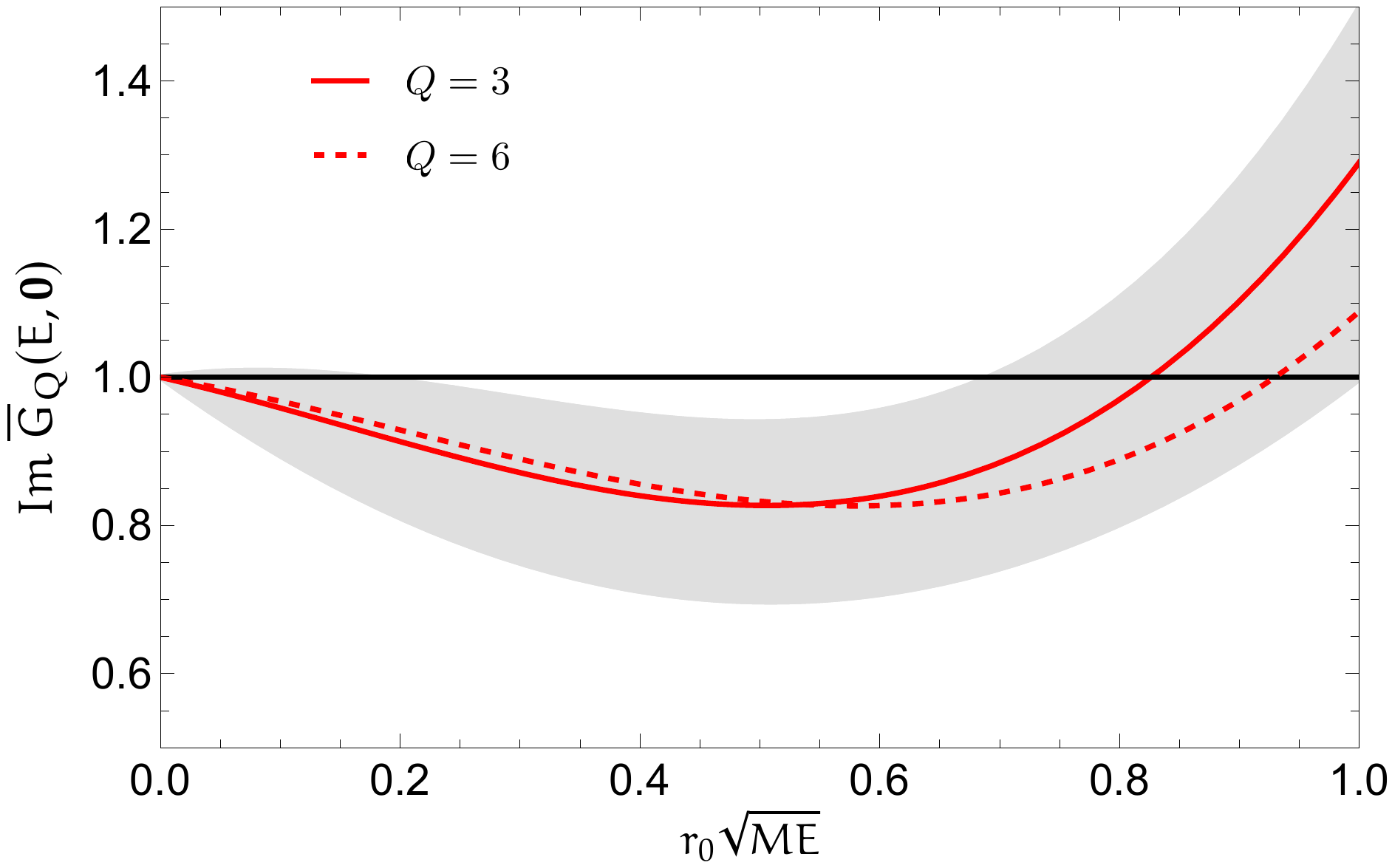}
\caption{Plot of the function $\Im{\overline G}_Q(E,{\mathbf 0})$ vs $r_0\sqrt{ME}$. The solid and dashed red
  curves corresponds to $Q=3$ and $Q=6$.
  The gray bands correspond to propagation of uncertainties in $h_{1,2}$.}  
  \label{fig:ImbarGr}
\end{figure}

Now, continuing back to Minkowski space, choosing the source points
$x_1=(t,{\mathbf x})$, $x_2=(0,{\mathbf 0})$, and using the Fourier
transform found in Appendix~\ref{app:ft} gives
\begin{multline}
{\cal C_Q}''(Q,E) =   \gamma^5 \pqty*{ - \frac{173\sqrt{2} \pi^2}{1920} \frac{h_1^2}{c_0} + \frac{7\sqrt{2} \pi^2}{64} h_2} \frac{\Gamma\left(\frac{3}{2}-\Delta_Q\right)}{\Gamma\left(\frac{5}{2}-\Delta_Q\right)} \\
\times\Bigg\lbrack
\psi\left(\frac{3}{2}-\Delta_Q\right)+\pi\tan\pi\Delta_Q -\log\left( E \lambda^{-1} \right)
\Bigg\rbrack .
\label{eq:r2effe}
\end{multline}
Note that
\begin{equation}
  \psi\left(\frac{3}{2}-\Delta_Q\right) \mapright{\Delta_Q \to \infty} -\frac{1}{\Delta_Q}\;+\; \order{\Delta_Q^{-2}}
  \;-\; \pi\tan\pi\Delta_Q\; +\;\log\left(\Delta_Q\right).
\end{equation}
Therefore, asymptotically, the harmonic contribution cancels, as is
also the case with the scattering length
corrections in Eq.~(\ref{eq:swscbe9})~\cite{Beane:2024kld}. The large logarithm can be removed
by now choosing $\lambda^{-1}= \Delta_Q M r_0^2$.  Plotting
$\Im{\overline G}_Q(E,{\mathbf 0})$ vs $r_0\sqrt{ME}$ in
Fig.~\ref{fig:ImbarGr} indicates the validity of the perturbative
expansion. One sees that there is very little variation between $Q=3$
and $Q=6$. Because effective-range effects enter as an irrelevant
deformation, perturbation theory works best for small effective ranges
and/or small energies. Note that the function ${\cal C_Q}''$ varies
more slowly with $Q$ than ${\cal C_Q}$ (asymptotically as $Q^{1/3}$ as
compared to $Q^{11/6}$).

\section{Unnuclear matter}
\label{sec:um}

\subsection{EFT of neutron matter}
\label{sec:eftns}

\begin{figure}[!ht]
\centering
\includegraphics[width = 0.9\textwidth]{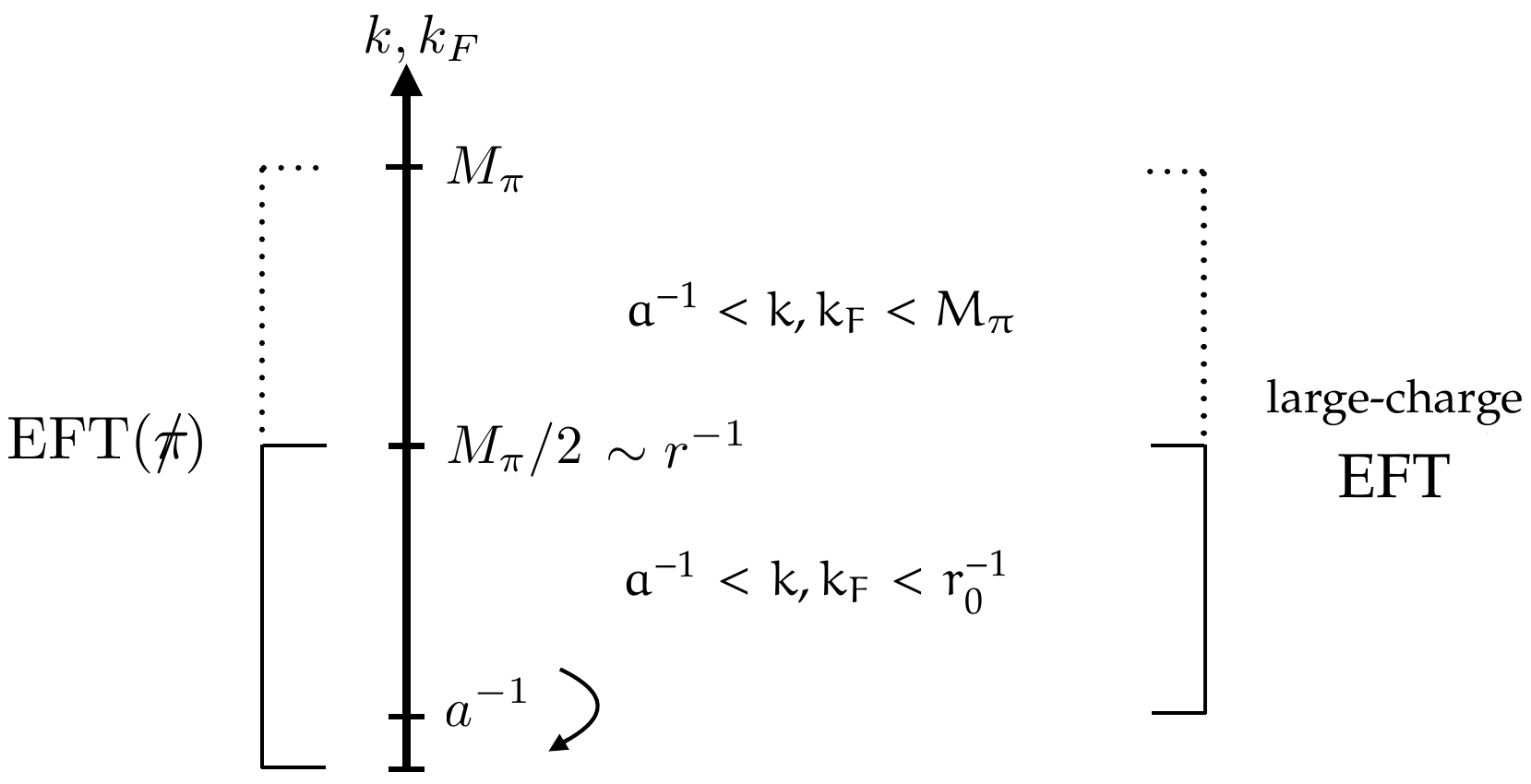}
\caption{Hierarchy of scales of $\nopi$. The dotted regions denote the ``effective'' region of validity
  of the \ac{eft}. The arrow denotes a transition from the large-charge \ac{eft}, which is defined about the nontrivial
unitary fixed point, to an \ac{eft} description about the trivial, non-interacting, fixed point.}
  \label{fig:eftpislash}
\end{figure}

\noindent To this point, the \ac{eft} of fermions near unitarity has been
treated as generic.  In this section, the focus will be on the specific
case of neutron matter. Following Ref.~\cite{Gandolfi:2022dlx}, the values of the
(unmeasured) neutron-neutron effective-range parameters are estimated
to be $a=-18.5$ fm and $r_0 = 2.7$ fm.  The neutrons have mass
$M=939$~MeV and the lightest \ac{dof} that is integrated out
of the \ac{eft} is the pion with mass $M_\pi = 140$~MeV, whose t-channel
exchange in the neutron-neutron scattering amplitude gives rise to a
branch-point singularity in the complex $k$-plane at $M_\pi/2$.  Hence
$r_0^{-1}\sim M_\pi/2$ formally sets the high-momentum scale in
$\nopi$, the \ac{eft} of contact operators which describes the interactions
of neutrons and protons at very low momentum transfers where the pion
is integrated out of the \ac{eft}. The branch-point singularity due to
t-channel pion exchange is in practice weak and therefore in most
applications $\nopi$ extends to $M_\pi$. While in $\nopi$ the low
scale is at zero, the threshold for scattering, in the large-charge
\ac{eft} the low scale is set by $a^{-1}$ as the \ac{eft} is an expansion about
the unitary fixed point. As one probes momenta $k$ such that $k a<1$
then a different \ac{eft} is required, which is an expansion about the
non-interacting fixed point. Therefore, the large-charge \ac{eft} of
neutron matter is formally valid for momenta in the range $a^{-1}<
k,\kf <r_0^{-1}$, and in practice may be valid in the range $a^{-1}<
k,\kf <M_\pi$.  Translating to energy scales, this corresponds to
energies between $0.1$ and $6$ MeV and $0.1$ and $21$ MeV,
respectively. The hierarchy of scales is illustrated in
Fig.~\ref{fig:eftpislash}.

\subsection{Deformation and perturbative window}
\label{sec:dandpw}

\begin{figure}[!ht]
\centering
\includegraphics[width = 0.9\textwidth]{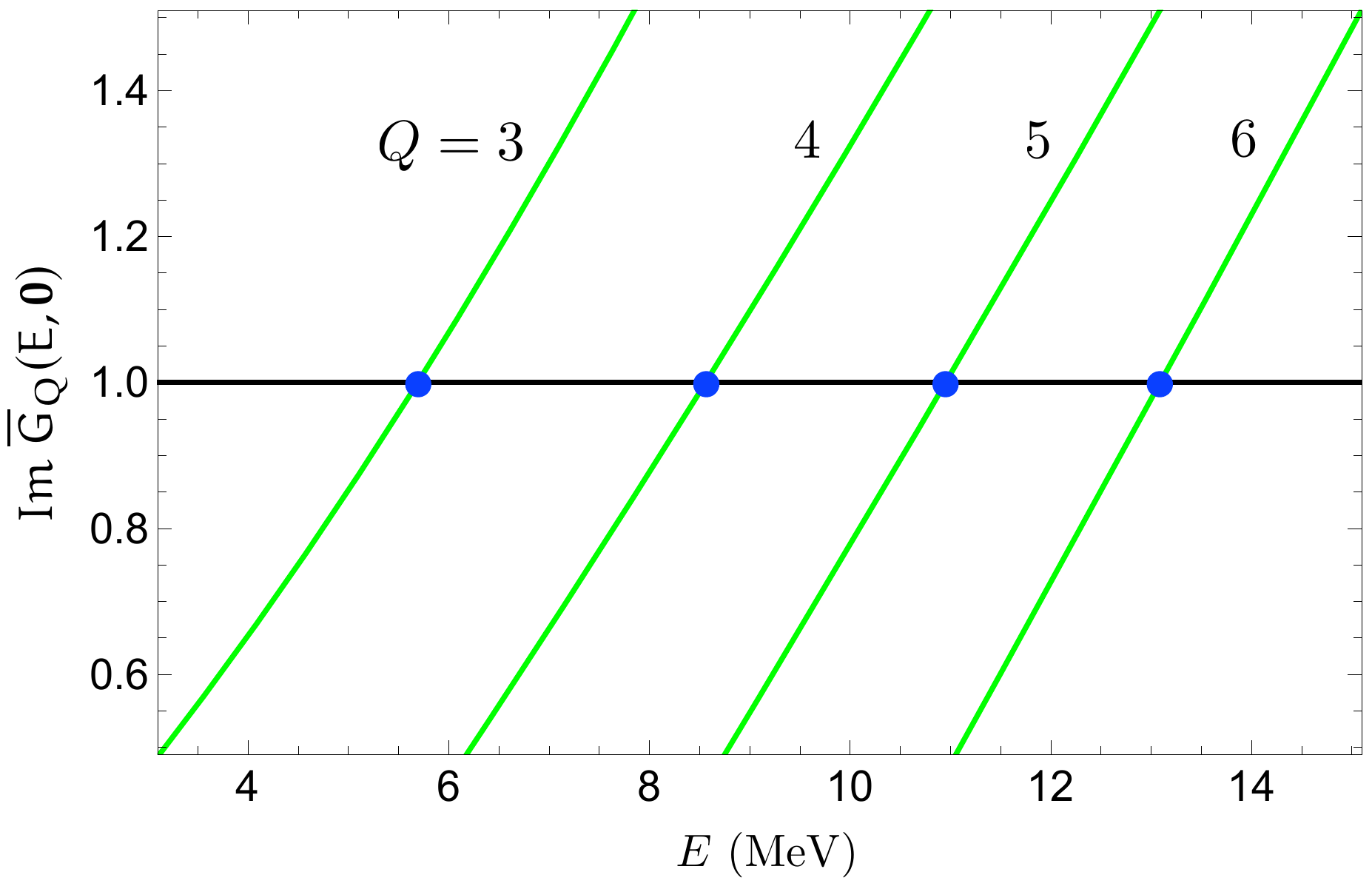}
\caption{Plot of the function $\Im{\overline G}_Q(E,{\mathbf 0})$ vs E (MeV). The green
  curves corresponds to $Q=3,4,5,6$. The blue dots denote the kinematical points at which the deformations cancel.}
  \label{fig:ImbarGunn}
\end{figure}

In neutron matter, both kinds of deformation should be considered and
therefore,
\begin{eqnarray}
\Im G(E,{\mathbf 0}) &=& C_0\;E^{\Delta_Q-5/2} \Big\lbrack 1  \ +\ {\cal C_Q}\,\left({a\,\sqrt{ME}}\right)^{-1} \ +\ {\cal C''_Q}\,{r_0^2\,{ME}}\Big\rbrack\; .
\end{eqnarray}
Fig.~\ref{fig:ImbarGunn} is a plot of $\Im{\overline G}_Q(E,{\mathbf
  0})$ vs E (MeV) for $Q=3,4,5,6$. Note that the substantial
uncertainties in the coupling constants have not been propagated in
this plot for clarity of presentation.  Perhaps the most interesting
observation is that the scattering-length and effective-range
deformations enter with opposite signs and there is therefore a
partial cancellation which implies that for each $Q$, there is an
energy at which the Schr\"odinger symmetry is restored. One sees that
the energy window in which perturbation theory is valid is rather
narrow for each $Q$ value.  However, as discussed in
the introduction, for instance at \ac{samu}, there is excellent energy
resolution of the final-state neutrons which in principle allows one
to focus on the regions of enhanced symmetry. The fundamental obstacle
from the theory side is the large uncertainties in the coupling
constants, particularly $h_2$. However, it may be possible to improve
these determinations using quantum \ac{mc} simulations of the energy
density.

\subsection{Three-body example}
\label{sec:tbe}

\begin{figure}[!ht]
\centering
\includegraphics[width = 0.88\textwidth]{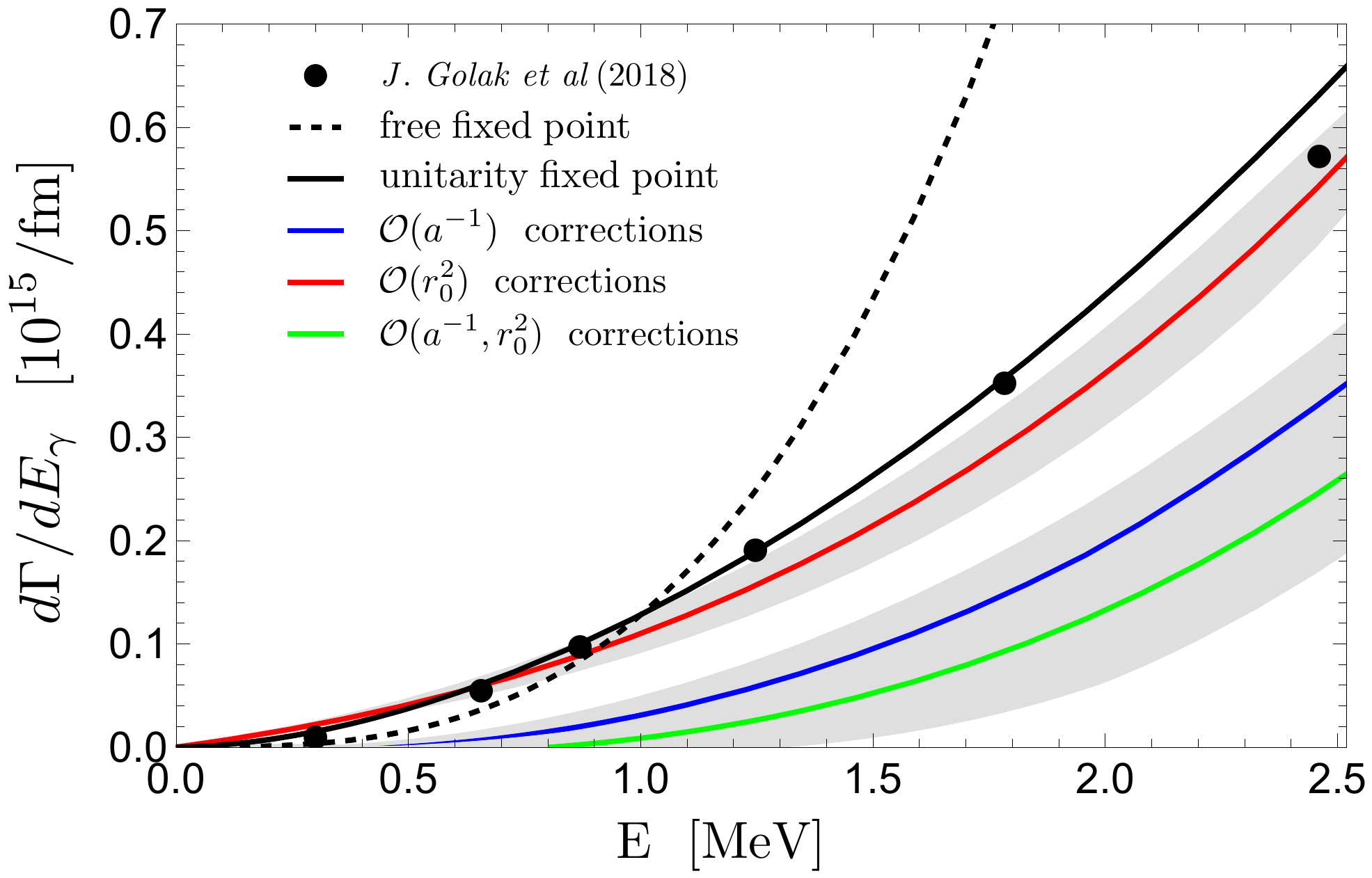}
\caption{Energy (center-of-mass) spectrum of three neutrons in the reaction ${}^3H(\pi^-,\gamma)3n$. The black circles give the calculations by Golak et al. in Ref.~\cite{Golak:2018jje}.
The various curves and regions are explained in the main text.}
  \label{fig:ImGtriton}
\end{figure}

Following Ref.~\cite{Hammer:2021zxb}, consider the process
\begin{equation}
	\pi^-\,+\,^3H\to \gamma\,+\, nnn,
\end{equation}
which has been studied in
Ref.~\cite{Golak:2018jje}, and whose theoretical predictions of the
capture rate, using the AV18 two-nucleon potential and the Urbana IX
three-body force, can be treated as ``data'' for the comparison with
the predictions from the large-charge \ac{eft} with perturbative
Schr\"odinger-symmetry breaking from scattering-length and
effective-range effects. The formula, Eq.~(\ref{eq:swscbe8c2}) is used
to obtain the capture rate, which is shown in
Fig.~\ref{fig:ImGtriton}. The dashed black curve follows from
considering the case of free neutrons, whose conformal dimension is
that of the three-body primary operator with fields assigned naive
dimensions, giving $11/2$. The black curve is the unitary fixed point
with the conformal dimension given by its well-known value for $Q=3$:
$4.27272$ (p-wave)~\cite{Chowdhury:2023ahp}~\footnote{Note that while
  the large-charge value of the conformal dimension at $Q=3$, taken from Eq.~\eqref{eq:cdLO},
  is only about one half this value, subleading corrections in the Schr\"odinger
  limit, which involve new undetermined constants~\cite{Son:2005rv}, can be chosen to reproduce
the $Q=3$ value exactly. This is done in Appendix~\ref{sec:icd}.}. As noted in
Ref.~\cite{Hammer:2021zxb}, the proximity of this curve to the data is
indicative of unnuclear behavior of the three-neutron final
state. While the effective-range corrections, given by the red curve,
appear controlled and perturbative, the scattering length corrections,
given by the blue curve, and the combined effect, given by the green
curve, appear to destroy this agreement. (Note that in all cases the
gray bands arise from the uncertainties in the Lagrange-density
parameters.) However, note from Fig.~\ref{fig:ImbarGunn}, that the
perturbative window is not approached for $Q=3$ until $E\sim 3$~MeV,
and indeed the green curve in Fig.~\ref{fig:ImGtriton} does cross
the solid black line at around $E\sim 6$~MeV, indicating a perturbative
window. Therefore the energies probed by this data are too low to offer a
meaningful test of the perturbative expansion. (Note as well that at very low energies
the blue curve is negative, in violation of unitarity.) Presumably the
perturbative window can be extended to lower energies by considering
$\order{1/a^2}$ corrections. A more ambitious proposal would be to sum
$1/a$ corrections to all orders in the two-point function of the
large-charge operator about the unitarity fixed point. This would give
an expression for the two-point function along the entire \ac{rg}
trajectory connecting the interacting and non-interacting fixed
points~\cite{Chowdhury:2023ahp}.

\section{Conclusion}
\label{sec:conc}

With few exceptions, quantum field theories of nature tend to be far
from conformal fixed points, and in most exceptions that are relevant
to experiment, require lattice field theory simulations in order to
extract physical observables.  An interesting system in this regard in
the non-relativistic domain is the unitary Fermi gas, which is
described by a \ac{nrcft}. This system is superfluid and remains
superfluid in the presence of small deformations away from the
Schr\"odinger symmetry limit. In the symmetry limit the superfluid \ac{eft}
has subsectors of fixed charge which admit a large-charge
expansion. This allows the exact closed-form determination of
correlation functions which encode the propagation of conformal
superfluid matter in space and time. In the nuclear physics context,
where, for instance, the unitary fermions are taken to be neutrons,
this conformal superfluid matter has been dubbed unnuclear
matter~\cite{Hammer:2021zxb}, as it does not have a particle
interpretation. Nuclear reactions with few-body systems of neutrons in
the final state offer a concrete experimental realization of unnuclear
matter. Of course, describing a few-body system of neutrons
as an expansion whose leading order is a \ac{nrcft} requires a quantitative
measure of the symmetry breaking due to scattering length, effective range
and other shape parameter effects which are clearly present in the two-neutron
system and explicitly break the Schr\"odinger symmetry. This has been done
here. The main conclusions of this study are:

\begin{itemize}

\item Schr\"odinger-symmetry breaking corrections to the large-charge
  two-point function have been computed in perturbation theory around the large-charge ground state. Closed-form
  expressions have been found for the Goldstone field perturbation by making use
  of a coordinate transformation to the oscillator frame, which decouples the
  temporal dependence.

\item The Schr\"odinger-symmetry breaking corrections are determined
  by a priori unknown Lagrange-density parameters.  These parameters
  contribute to the energy density of the superfluid matter, and have
  been computed using quantum \ac{mc} simulations. The predictive power of the large-charge \ac{eft}
  relies on the accuracy with which these parameters are determined.

\item The $\order{r_0}$ effective-range effects have been found to
  vanish in the large-charge \ac{eft}, as was found for the three-body case
  in Ref.~\cite{Chowdhury:2023ahp}.  The $\order{r^2_0}$
  effective-range effects have been calculated, and together with the
  $\order{a^{-1}}$ scattering-length effects calculated in
  Ref.~\cite{Beane:2024kld}, provide the leading
  Schr\"odinger-symmetry breaking corrections to the large-charge
  two-point function. Critically, the two kinds of deformation enter
  with opposite signs and therefore there is a partial cancellation.

\item At fixed charge $Q$, the energy-dependent Schr\"odinger-symmetry
  breaking corrections to the large-charge two-point function are
  found to be perturbative over a range of (center-of-mass) energies
  of the $Q$-neutron system that decreases with increasing $Q$. These results
  provide a guide to the energy regions that experimentalists could probe which
  allow a controlled \ac{eft} description.  

\end{itemize}

As regards future work, it would be interesting to study higher-order
terms including the contributions from the droplet
edge~\cite{Hellerman:2020eff,Hellerman:2021qzz}, which are expected to
scale with fractional powers of the deformation parameters.  One could
furthermore consider the perturbation theory relevant to
Schr\"odinger-symmetry breaking for the large-charge expansion in the
case of two spatial dimensions. In addition, while the focus of this
paper has been on the large-charge two-point function,
symmetry-breaking effects in three- and higher-point functions may
also be of interest experimentally.

  \section*{Acknowledgments}

  \begin{small}\sffamily
The authors would like to thank S.~Aoki, J.A.~Carlson, S.~Gandolfi,
H.W.~Hammer, S.~Hellerman, D.B.~Kaplan, D.R.~Phillips, and D.T.~Son
for useful discussions.  This work was supported by the Swiss National
Science Foundation under grant number 200021\_219267. In addition,
S.R.B is supported by the U.~S.~Department of Energy grant
\textbf{DE-FG02-97ER-41014} (UW Nuclear Theory).  D.O. and
S.R. gratefully acknowledge support from the Yukawa Institute for
Theoretical Physics at Kyoto University, as well as the from the
Simons Center for Geometry and Physics at Stony Brook University,
where some the research for this paper was performed.
  \end{small}

\newpage

\appendix
\section{Improved conformal dimension}
\label{sec:icd}

\begin{figure}
\centering
\includegraphics[width = 0.85\textwidth]{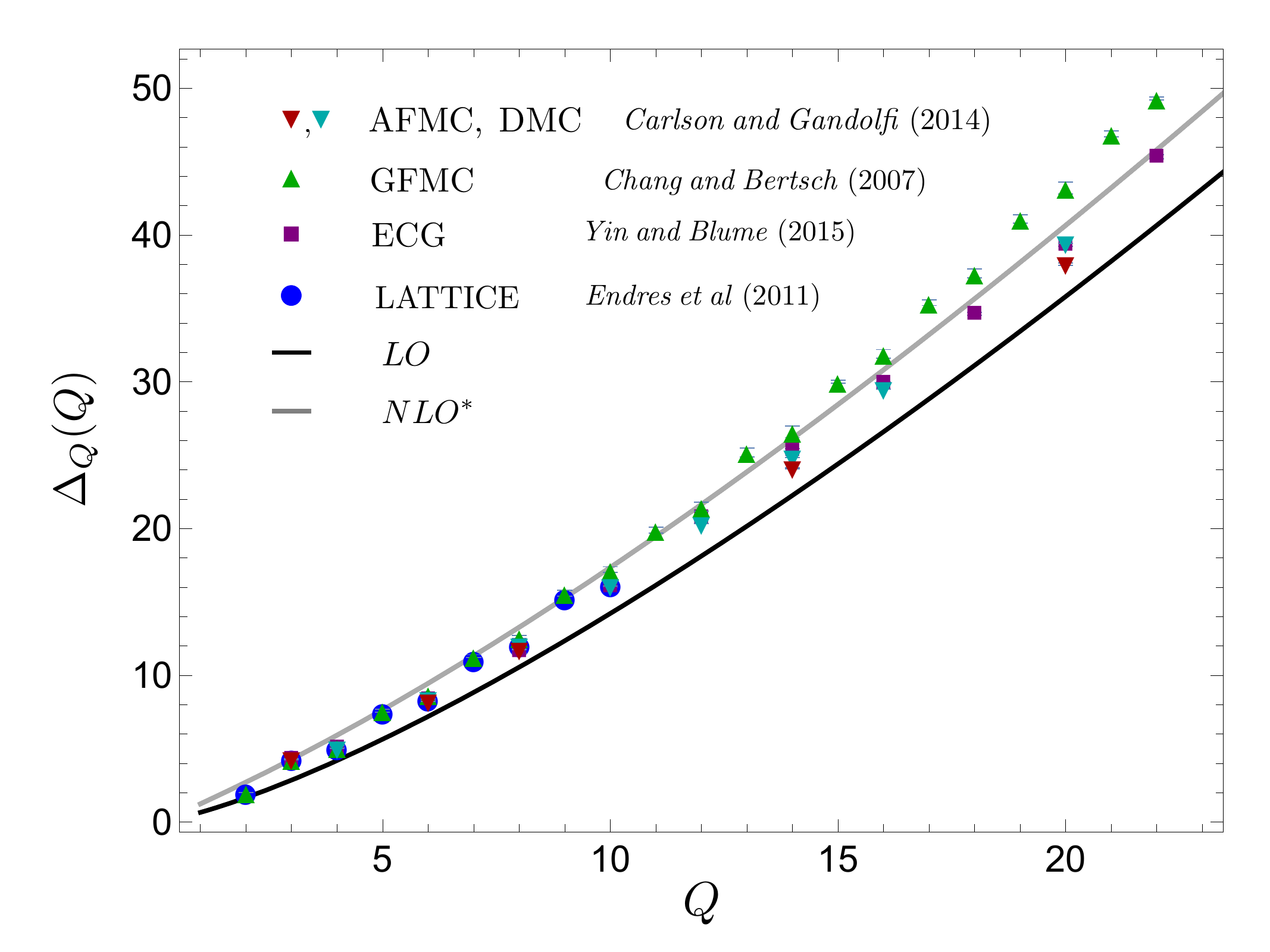}
\caption{Fit of DMC simulation data up to $Q=20$, as described in the
  text. The solid black line is leading order in the large-charge
  expansion, and the gray band is the \ac{nlo} fit, including the Casimir
  correction. The simulation data are as described in the text.}
  \label{fig:delqvsqfit}
\end{figure}

When considering applications of the large-charge \ac{eft} at low-$Q$
values, say $Q=3-6$, it is sensible to include higher-order
corrections~\cite{Son:2005rv,Favrod:2018xov,Kravec:2019djc,Hellerman:2021qzz,Beane:2024kld}. In
the Schr\"odinger-symmetry limit, the large-charge conformal dimension
to \ac{nlo} takes the form
\begin{eqnarray}
\Delta_Q(Q) \; =\;  \frac{3^{4/3}}{4} \xi^{1/2}Q^{4/3}\;-\; 3^{2/3 }\sqrt{2}\pi^2 \xi\,c_{\textsc{NLO}}\, Q^{2/3} \; +\; \order{Q^{5/9}} \; +\; \ldots \; +\; \frac{1}{3\sqrt{3}}\log Q \ ,
\label{eq:icd}
\end{eqnarray}
where, in addition, the universal Casimir correction has been
included~\cite{Hellerman:2021qzz}.  The coefficient of the \ac{nlo}
term may be fit to simulation data. Fig.~\ref{fig:delqvsqfit} shows
simulation data up to $Q=22$ from Green's function \textsc{mc} (\textsc{gfmc})~\cite{Chang:2007zzd},
lattice \ac{mc}~\cite{Endres:2011er}, both diffusion \textsc{mc} (\textsc{dmc}) and auxiliary field \textsc{mc} (\textsc{afmc})~\cite{Carlson:2014pxa},
and a correlated Gaussian basis set expansion (ECG)~\cite{Yin_2015}.
Fitting to the \textsc{afmc} data over the range $Q=3-20$ one finds $c_{\textsc{nlo}}=-0.053715(1)$.
This fit is illustrated in Fig.~\ref{fig:delqvsqfit}.

\section{Fourier transforms}
\label{app:ft}

The following Fourier transforms~\cite{Hammer:2021zxb,Braaten:2023acw} are useful:
\begin{multline}
  \int dt \int d^3{\mathbf x}\; \theta(t)\,t^{-\Delta}\,\exp\left({i\frac{Q M \mathbf{x}^2}{2 t}}\right) \exp\left( i E t - i{\mathbf p}\cdot{\mathbf x}\right)\\
 =  i^{\Delta-1} \left( \frac{2\pi}{QM}\right)^{3/2}   \left(\frac{p^2}{2QM}-E\right)^{\Delta-5/2}\Gamma\left(\frac{5}{2}-\Delta\right) \ ,
\label{eq:swscbe6}
\end{multline}
and 
\begin{multline}
 \int dt \int d^3{\mathbf x}\; \theta(t)\,t^{-\Delta}\,\log\left(i \lambda t\right)\exp\left({i\frac{Q M \mathbf{x}^2}{2 t}}\right) \exp\left( i E t - i{\mathbf p}\cdot{\mathbf x}\right) \\
  =  i^{\Delta-1} \left( \frac{2\pi}{QM}\right)^{3/2}   \left(\frac{p^2}{2QM}-E\right)^{\Delta-5/2}\Gamma\left(\frac{5}{2}-\Delta\right) \\
  \times \bqty*{\psi\left(\frac{5}{2}-\Delta\right) - \log*( \frac{1}{\lambda} \left(\frac{p^2}{2QM}-E\right) ) }\ .
\label{eq:swscbe7}
\end{multline}

\section{Variation of parameters}
\label{sec:variation-parameters}

The differential equations studied in this paper reduce to (non-homogeneous) second-order \acp{ode} of the form
\begin{equation}
  u''(x) + p(x) u'(x) + q(x) u(x) = f(x) .
\end{equation}
There is a general solution of this equation, which is obtained using the \emph{variation of parameters}~\cite{Lagrange}.
Let \(u_1(x)\) and \(u_2(x)\) be independent solutions to the associated homogeneous problem, \emph{i.e.} the one with \(f(x) = 0\):
\begin{equation}
  u_{1,2}''(x) + p(x) u_{1,2}'(x) + q(x) u_{1,2}(x) = 0 .
\end{equation}
Then the general solution to the non-homogeneous problem is
\begin{equation}
  u_G(x) = A(x) u_1(x) + B(x) u_2(x)  ,
\end{equation}
where
\begin{align}
  A(x) &= - \int^x \dd{\xi} \frac{u_2(\xi) f(\xi)}{W(\xi)} &   B(x) &= \int^x \dd{\xi} \frac{u_1(\xi) f(\xi)}{W(\xi)} ,
\end{align}
and \(W(x)\) is the Wronskian of \(u_1(x)\) and \(u_2(x)\):
\begin{equation}
  W(x) = u_1(x) u_2'(x) - u_1'(x) u_2(x) .
\end{equation}

\section{The boundary term}\label{sec:boundaryTerm}

In the usual case it is assumed that fields vanish at spatial and
temporal infinity.  One consequence of this is that in perturbation
theory, the solution at order \(n\) allows one to compute the action
at the saddle at order \(n + 1\), since there are no boundary terms.
This is not the case for the action in the oscillator frame, because
the couplings are time-dependent and grow exponentially at \(\tilde
\tau = \pm \infty\).

As a simple example of such a system, consider the Lagrangian
\begin{equation}
	L = \frac{1}{2} \dot \phi^2 + r_0 \cosh(\omega \tilde{\tau}) F(\dot\phi)
\end{equation}
for $r_0 \ll 1$.

The \ac{eom} is 
\begin{align}
	0= \del_t \frac{\delta L}{\delta \dot \phi} &= \del_{\tilde{\tau}} \left[ \dot \phi + r_0 \cosh(\omega \tilde{\tau}) F'(\dot \phi) \right]\\
	&= \ddot \phi + r_0 \cosh(\omega \tilde{\tau}) \ddot\phi F'(\dot\phi) + r_0 \omega \sinh(\omega \tilde{\tau}) F'(\dot\phi).
\end{align}
The solution may be written as
\begin{equation}
	\phi = \phi_0 + r_0 \phi_1,
\end{equation}
so the \ac{eom} at first order takes the form
\begin{equation}
	\ddot \phi_0 + r_0 \ddot\phi_1 + r_0 \cosh(\omega \tilde{\tau}) \ddot\phi_0 F'(\dot\phi_0) + r_0 \omega \sinh(\omega \tilde{\tau}) F'(\dot\phi_0) = 0 ,
\end{equation}
and, order by order,
\begin{align}
  \phi_0&= A \tilde{\tau} + B, &  \phi_1&= C \tilde{\tau} + D - \frac{F'(A)}{\omega} \sinh(\omega \tilde{\tau}).
\end{align}
Note that $\phi_1$ does not vanish at the boundary $t=\pm T$.

The action at the saddle thus receives three contributions:
\begin{enumerate}
	\item the \ac{lo} action evaluated on the \ac{lo} solution
	\item the \ac{lo} action evaluated on the \ac{nlo} solution
	\item the \ac{nlo} action evaluated on the \ac{lo} solution.
\end{enumerate}
Contribution 2 does not vanish at the saddle due to a boundary contribution given by
\begin{equation}
	\phi_1 \frac{\delta L}{\delta \dot \phi}\Bigg|_{-T}^T = \left( 2CT - \frac{2F'(A)}{\omega}\sinh(\omega T) A \right) .
\end{equation}
Explicitly, the action at the saddle at order \(\order{r_0}\) reads:
\begin{equation}
  \begin{aligned}
    S[\phi_0+r_0\phi_1] &= \int_{-T}^T \dd{\tilde{\tau}} \frac{1}{2}\dot \phi_0^2 + \dot\phi_0\dot \phi_1 r_0 + \cosh(\omega t) F(\dot\phi_0)r_0  \\
                  &= \int_{-T}^T \dd{t} \frac{1}{2} A^2 + A(C-\cosh(\omega t) F'(A))r_0 +  \cosh(\omega t) F(A)r_0 \\
                  &= A^2T+ \left[2ACT - \frac{2AF'(A)}{\omega} \sinh(\omega T) \right]r_0 + \frac{2F(A)}{\omega}\sinh(\omega T)r_0 ,
  \end{aligned}
\end{equation}
which depends manifestly on the \(\order{r_0}\) value of the field \(\phi_{1}\).

\section{Structure of the solution to the bulk EFT}
\label{sec:struct-solution-bulk}

It is informative to consider the general structure of the solution to the \ac{eom}, in the presence of effective-range corrections
and the corresponding expansion of the action at the saddle. The Lagrange density in the oscillator frame takes the form 
\begin{equation}
	\tilde {\cal L} = - c_0 \tilde X^{5/2} + \frac{\mu}{\omega} \sum_{k=1} h_k r_0^k \tilde X^{(5+k)/2} \mu^{k/2}\cosh^k(\omega \tilde\tau).
\end{equation}
Consider a solution to the \ac{eom} of the form
\begin{equation}
	\tilde \theta(\tilde \tau, \tilde v) = \tilde \theta_0(\tilde \tau, \tilde v) + \frac{\mu}{\omega} \sum_{k=1} (r_0\mu^{1/2})^k \tilde \theta_k(\tilde \tau, \tilde v),
\end{equation}
which is an expansion around the solution $\tilde \theta_0(\tilde
\tau, \tilde v) = -i\mu\tau$ of the undeformed Schr\"odinger
problem. The \ac{eom} turns into a hierarchy for the $\tilde\theta_k$, where at each order there is an inhomogeneous \ac{pde}
with a source of the form
\begin{equation}
	f_k(\tilde\tau, \tilde v) = \varphi_k(\tilde v) \frac{\dd{}}{\dd{\tilde\tau}}\cosh^k(\omega \tilde\tau),
\end{equation}
where the $\varphi_k$ depend on the solutions at order $k'<k$:
\begin{align}
	F_k(\tilde{\ddot \theta}_k, {\tilde\theta}''_k, \tilde\theta'_k|\tilde \theta_{k'}) &= \varphi_k(\tilde v) \frac{\dd{}}{\dd{\tilde\tau}}\cosh^k(\omega \tilde\tau), & k'<k,
\end{align}
and where $F_k$ is by construction linear in $\tilde\theta_k$.

Using the fact that the \ac{rhs} can be decomposed into a sum of hyperbolic sine functions,
\begin{equation}
  \odv{\cosh^k(\omega \tilde{\tau})}{\tilde{\tau}} = \frac{\omega}{2^k} \sum_{l=0}^{\floor{k/2}} \binom{k}{l} \pqty{k - 2l} \sinh*(( k - 2l) \omega \tilde{\tau}),
\end{equation}
the variables can be separated by writing
\begin{equation}
  \tilde{\theta}(\tilde{\tau}, \tilde{v}) = \sum_{l=0}^{\floor{k/2}} \sinh*( (k - 2l) \omega \tilde{\tau}) B_{k,l}(\tilde{v}),
\end{equation}
so that the \ac{eom} turn into a hierarchy of inhomogeneous \ac{ode} for the functions \(B_{k,l}\).

The transformation to the flat frame is obtained by using the identity
\begin{equation}\label{eq:sinh-arctanh}
  \sinh*( n \arctanh(\omega \tau)) = \frac{1}{\pqty*{ 1 - \omega^2 \tau^2}^n} \sum_{k=0}^{\floor{n/2}} \binom{n}{2k + 1} \pqty{ \tau \omega}^{2k + 1} = \frac{P_n(\omega \tau)}{\pqty*{ 1 - \omega^2 \tau^2}^n}.
\end{equation}
In what follows one can rename
\begin{equation}
  \omega \tau = z  
\end{equation}
in order to disentangle the parametric dependence of all the terms.
Then,
\begin{equation}
  \theta_k(z, v) = \sum_{l=0}^{\floor{k/2}} \frac{P_{k-2l}(z)}{\pqty*{ 1 - z^2}^{k/2 - l}} B_{k,l}(v),
\end{equation}
and it follows that \(X_k\) takes the form
\begin{equation}
  X_k(\tau, v) = \mu \sum_{l=0}^{\floor{k/2}} \frac{Q_{k,l}(z^2, v)}{\pqty*{ 1 - z^2}^{k/2 + 1 - l}} ,
\end{equation}
where \(Q(z^2, v)\) is a polynomial in \(z^2\).

The analytic structure in the \(\tau\) plane around the points \(\pm
1/\omega\), that is \(z = \pm 1\), is of special interest as this
determines which terms will contribute to the final result.  In the
perturbative expansion of \(X\), each term of order \(\order{r_0^k}\)
contains singularities that are either poles if \(k\) is even, or 
branch cuts if \(k\) is odd.  The same structure, with the roles of
even and odd interchanged is found in the Lagrange density evaluated at the
saddle,
\begin{equation}
  {\cal L}_{\saddle} = - c_0 \frac{\mu^{5/2} L_0}{\pqty*{ 1 - z^2}^{5/2}} - \mu^{5/2} \sum_k (r_0 \sqrt{\mu})^k \sum_{l=0}^{\floor{k/2}} \frac{L_{k,l}(z^2, v)}{\pqty*{ 1 - z^2}^{(k+5)/2 - l}} .
\end{equation}
From Eq.~\eqref{eq:flat-frame-measure} it is clear that the integration measure in terms of \(\psi\) is given by
\begin{equation}
  \dd{\tau} \dd^3{x} = \frac{1}{\omega} R_{\ac{lo}}^3 \pqty{1 - z^2}^{3/2} \sin^2(\psi) \cos(\psi) \dd{\psi} \dd{z} \dd{\Omega} \, .
\end{equation}
It follows that the role of poles and branch cuts is again interchanged, and
\begin{multline}
  S_{\saddle} = - \frac{5 \pi^2 c_0 \mu^4}{16 \sqrt{2} \omega^4} \int_{-1}^1 \frac{\dd{z}}{1 - z^2} \\ - \frac{8 \pi \sqrt{2} \mu^4  }{\omega^4} \sum_k (r_0 \sqrt{\mu})^k \sum_{l=0}^{\floor{k/2}} \int_{-1}^1 \frac{\dd{z}}{\pqty*{ 1 - z^2}^{k/2 + 1 - l}} \\ \int \sin^2(\psi) \cos(\psi) \dd{\psi} L_{k,l}( z^2, \cos(\psi)).
\end{multline}

The integral over \(\psi\) gives a contribution that is a polynomial in \(z^2\), but it is still necessary to regularize the integral over \(z\).
In general these integrals will take the form
\begin{equation}
  \mathscr{I}(n,m) = \int_{-1}^1 \dd{z} \frac{z^{2m}}{\pqty*{1 - z^2}^n} ,
\end{equation}
where \(n\) is either integer (if \(k\) is even) or half-integer (if \(k\) is odd), and \(m = 0, 1, \dots, \floor{n}\).

One possible way of regularizing the \(\mathscr{I}(n,m)\) is to use analytic continuation in terms of gamma functions:
\begin{equation}
  \mathscr{I}(n,m) = \lim_{\delta \to 0} \mathscr{I}(n+\delta, m) = \lim_{\delta \to 0} \frac{\Gamma(m + \sfrac{1}{2}) \Gamma(1 - n - \delta)}{\Gamma(\sfrac{3}{2} + m - n - \delta)} .
\end{equation}
The gamma function has no zeros, but it has simple poles at negative integers.
It follows that there are two possibilities:
\begin{itemize}
\item If \(n\) is half-integer (\(k\) is odd), the denominator has a pole for \(\delta \to 0\) and the integral vanishes
  \begin{equation}
    \mathscr{I}(\setZ + \sfrac{1}{2}, m) = 0 \, .
  \end{equation}
\item If \(n\) is an integer (\(k\) is even), the integral has a pole:
  \begin{equation}
    \mathscr{I}(n+\delta, m) = \frac{(-1)^n \Gamma(m + \sfrac{1}{2})}{\Gamma(\sfrac{3}{2} + m - n) \Gamma(n)} \frac{1}{\delta} + \order{\delta^0}.
  \end{equation}
\end{itemize}

An alternative regularization is obtained by shifting the boundaries of integration:
\begin{equation}
  \mathscr{I}(n,m) = \lim_{\epsilon \to 0} \int_{-1+\epsilon}^{1-\epsilon} \dd{z} \frac{z^{2m}}{\pqty*{1 - z^2}^n} .
\end{equation}
For \(n \) integer the integral has a logarithmic divergence in \(\epsilon \to 0\), whose coefficient is precisely the same as the residue of the pole in \(\delta \to 0\):
\begin{equation}
   \int_{-1+\epsilon}^{1-\epsilon} \dd{z} \frac{z^{2m}}{\pqty*{1 - z^2}^n} = \frac{(-1)^n \Gamma(m + \sfrac{1}{2})}{\Gamma(\sfrac{3}{2} + m - n) \Gamma(n)} \log(\epsilon) + \order*{\frac{1}{\epsilon^{n - m - 1}}} + \order{\epsilon^0} .
\end{equation}

The finite terms \(\order{\delta^0}\) and \(\order{\epsilon^0}\) are
different.  This scheme dependence is however not a problem since it
simply corresponds to different normalizations of the operators in the
two-point function.  The only physically-relevant parameter is the
coefficient of the logarithm, which is determined unambiguously.

This structure holds for the bulk terms. However, there are
contributions due to the fact that the density decreases sharply at
the droplet edge. There are thus boundary effects that can lead to new
corrections that scale differently.  In the study of the \ac{nlo}
corrections (Section~\ref{sec:first-order-range}), it was found that
the size of the droplet is reduced by an amount proportional to
\(r_0^{2/3}\), which translates into a correction of
\(\order{r_0^{7/3}}\).
The appearance of new
fractional powers (and logarithmic terms) has already been observed in
conjunction with boundary effects in the large-charge expansion in the
undeformed Schrödinger-symmetric
case~\cite{Hellerman:2020eff,Hellerman:2021qzz}.

Up to this point the results have been expressed as functions of the
parameter \(\mu\). In Section~\ref{sec:continuity-charge}
the relationship between \(\mu\) and \(Q\) was found to be independent of
\(r_0\) at \ac{nlo} because of the cancellation of all the
\(\order{r_0}\) terms.  However, the same cancellation must occur at
all orders.  Each power of \(r_0\) in the expression of the density
\(\rho\) is accompanied by powers of \(\pqty{1 - \omega^2 \tau^2}\).
All these terms have to cancel separately already on the \ac{lhs} of
the differential form of the continuity
equation~(\ref{eq:continuity}), because there is no such \(\tau\)
dependence on the \ac{rhs}.  The conclusion is therefore that the expression of
\(\mu\) as function of \(Q\) in Eq.~(\ref{eq:mu-and-Q}) is valid at
all orders in \(r_0\): \(\pqty*{\mu/\omega}^3 = 3 \xi^{3/2} Q\).

\setstretch{1}

\printbibliography{}

\end{document}